\documentclass[aps,prl,twocolumn,superscriptaddress,floatfix]{revtex4-1}
\usepackage[utf8]{inputenc}
\usepackage[english]{babel}
\usepackage{amsmath}
\usepackage{feynmf}
\usepackage{xcolor}
\usepackage{soul}
\usepackage{graphicx}
\usepackage{amsfonts}
\usepackage{blindtext}
\usepackage{appendix}
\usepackage[colorlinks = true,
            linkcolor = blue,
            urlcolor  = blue,
            citecolor = blue,
            anchorcolor = blue]{hyperref}\usepackage{siunitx}\usepackage{siunitx}
\usepackage{siunitx}
\usepackage{bbold}

\usepackage{soul}

\usepackage[normalem]{ulem}

\begin{abstract}
Bosonic codes allow the encoding of a logical qubit in a single component device, utilizing the infinitely large Hilbert space of a harmonic oscillator. In particular, the Gottesman-Kitaev-Preskill code has recently been demonstrated to be correctable well beyond the break-even point of the best passive encoding in the same system. Current approaches to quantum error correction (QEC) for this system are based on protocols that use feedback, but the response is based only on the latest measurement outcome. In our work, we use the recently proposed Feedback-GRAPE (Gradient Ascent Pulse Engineering with Feedback) method to train a recurrent neural network that provides a QEC scheme based on memory, responding in a non-Markovian way to the full history of previous measurement outcomes, optimizing all subsequent unitary operations. This approach significantly outperforms current strategies and paves the way for more powerful measurement-based QEC protocols.
\end{abstract}

\begin{document}

\title{Non-Markovian feedback for optimized quantum error correction}

\author{Matteo Puviani}
\email{matteo.puviani@mpl.mpg.de}
\affiliation{Max Planck Institute for the Science of Light, 91058 Erlangen, Germany}

\author{Sangkha Borah}
% \email{}
\affiliation{Max Planck Institute for the Science of Light, 91058 Erlangen, Germany}
\affiliation{Department of Physics, Friedrich-Alexander Universität Erlangen-Nürnberg, 91058 Erlangen, Germany}

\author{Remmy Zen}
% \email{}
\affiliation{Max Planck Institute for the Science of Light, 91058 Erlangen, Germany}

\author{Jan Olle}
% \email{}
\affiliation{Max Planck Institute for the Science of Light, 91058 Erlangen, Germany}

\author{Florian Marquardt}
\email{florian.marquardt@mpl.mpg.de}
\affiliation{Max Planck Institute for the Science of Light, 91058 Erlangen, Germany}
\affiliation{Department of Physics, Friedrich-Alexander Universität Erlangen-Nürnberg, 91058 Erlangen, Germany}

\maketitle

\textit{Introduction}. 
Qubits, the key components of quantum computers, can be implemented in a variety of hardware platforms, such as superconducting circuits  \cite{devoret2004superconducting, CampagneIbarcq2020,Sivak2023}, trapped ions \cite{Kienzler2017, Fluehmann2018, Flhmann2019}, Rydberg atoms \cite{Wu2021}, and photonic systems \cite{konno2023propagating}. Irrespective of the platform, quantum decoherence that arises from relaxation and dephasing represents the greatest limitation for building quantum computers that are able to run large scale quantum algorithms.
The main objective of current research is the implementation of reliable \textit{quantum error correction} (QEC). Reliable QEC protocols can correct errors on logical qubits, extending the lifetime of the logical information stored in them far beyond the coherence time of individual physical components used for encoding a logical qubit. This will be useful for the purposes of quantum memories, quantum communication and quantum computing. \\
\begin{figure}[hb!]
\centering
\includegraphics[width=1.0\linewidth]{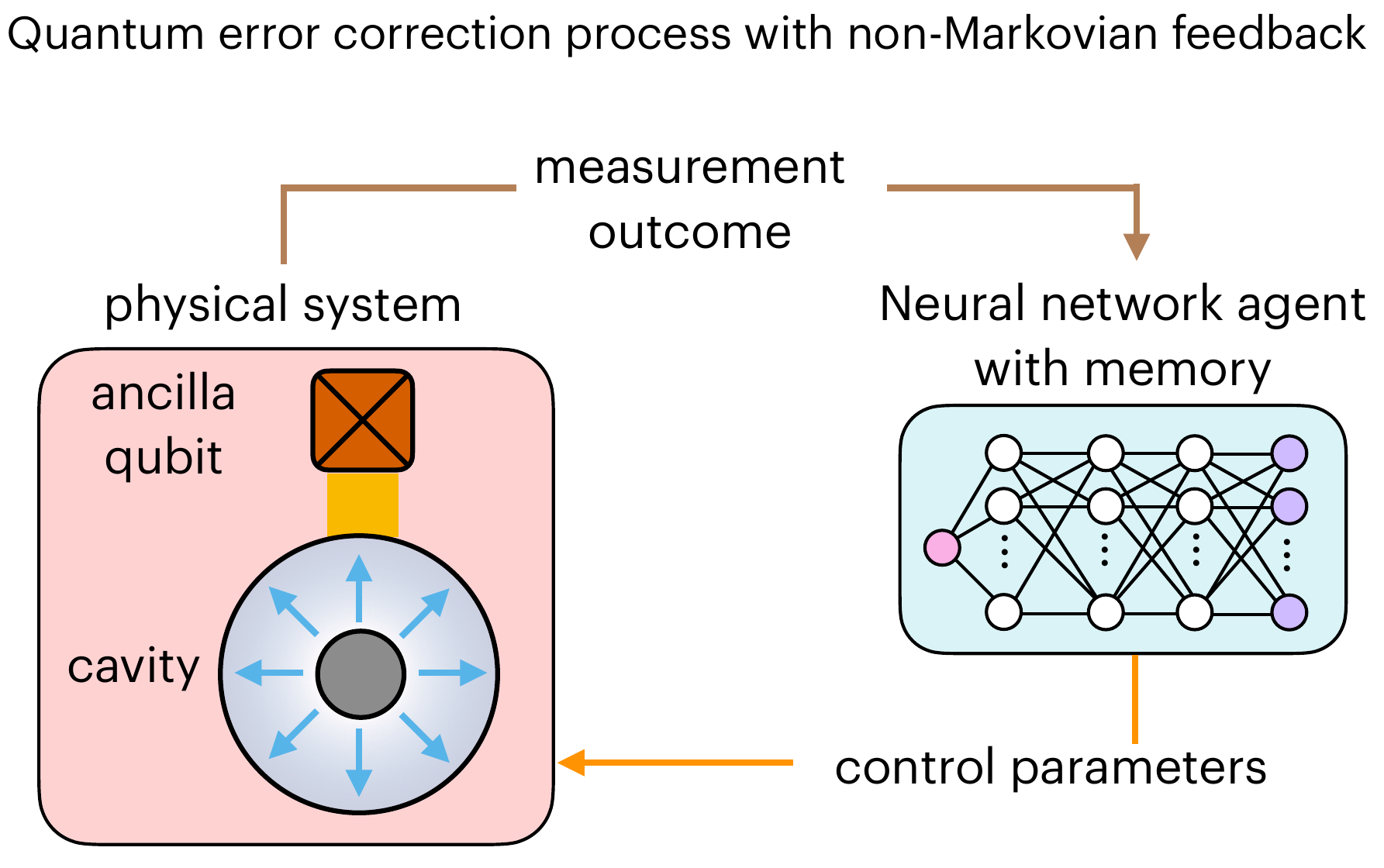}
\caption{Scheme of the proposed quantum error correction process via non-Markovian feedback. The physical system is represented by a cavity coupled to a control qubit. The result of each binary quantum measurement is used as input of a neural network with memory, which responds to all the sequence of measurements providing the optimized parameters to control the system.}
\label{scheme}
\end{figure}
\begin{figure*}[t!]
\centering
\includegraphics[width=1.0\linewidth]{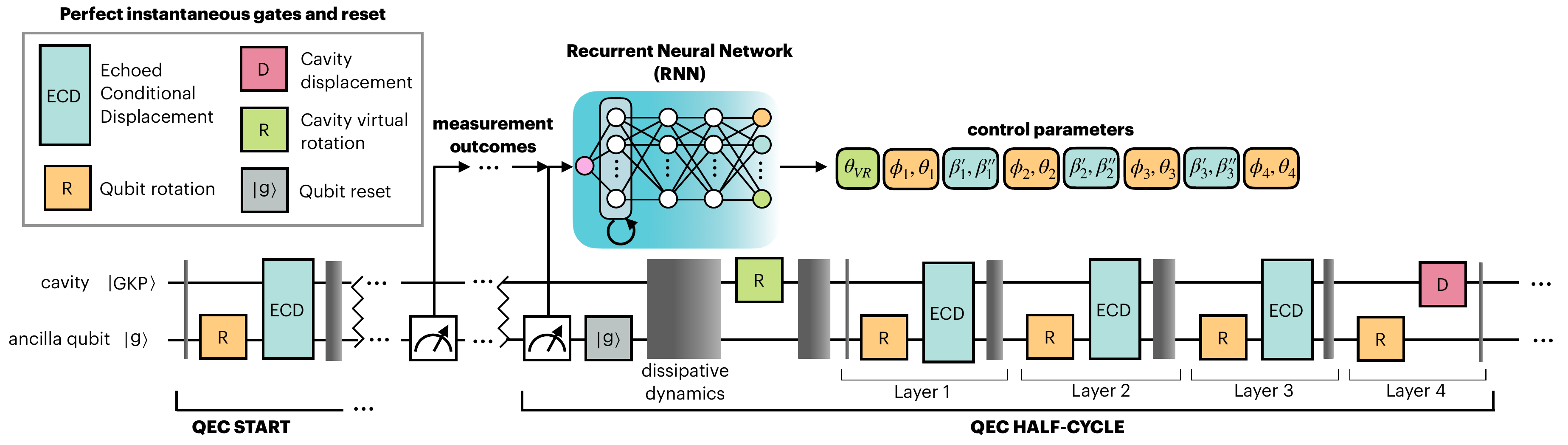}
\caption{Quantum error correction feedback protocol. The scheme represents the sBs QEC protocol with the NMF approach implemented with a RNN. On each cycle, the latest outcome of the measurement performed on the qubit is used as input for the RNN, which keeps memory of all the previous sequence of inputs and suggests the parameters used in the subsequent gates. The grey areas in the QEC circuit represent the duration of the dissipative dynamics \cite{supplement}.} \label{scheme_QEC}
\end{figure*} 
\indent The most widespread class of approaches to QEC is represented by the so-called \textit{stabilizer codes}, where $n$ physical qubits are used to encode $k < n$ logical qubits  \cite{Shor1995, Shor1996, Steane1996, FT2007}. This redundancy, along with the concept of quantum entanglement, can be used to correct a limited number of independent physical errors \cite{gottesman1997stabilizer}. An alternative to such qubit-based encoding is to use systems with continuous variables, namely bosonic systems, leading to \textit{bosonic encodings} \cite{Cai2021,xu2023faulttolerant}. Unlike the qubit-based encoding, the redundancy is achieved within a single physical component by exploiting the infinitely large Hilbert space of bosonic systems (i.e. quantum oscillators, such as photons in an optical cavity) \cite{Puri2017} to be able to perform QEC. Bosonic codes are more hardware-efficient as there is often only one or very few dominant decay channels to be corrected for, and they offer exceedingly long coherence times in the microwave range. However, they need nonlinear elements for processing, they require a protocol to prepare states and the overall strategies are still less developed than for multi-qubit encoding \cite{Puri2021}. \\
\indent A variety of bosonic encodings exist, such as Pegg-Barnett \cite{Barnett1986}, cat and kitten codes \cite{Cochrane1999, Mirrahimi2014, Ofek2016, Lescanne2020, Grimm2020,Xu2023}, \textit{Gottesman-Kitaev-Preskill} (GKP) \cite{GKP2001, PhysRevLett.125.260509, CampagneIbarcq2020, Puri2021, brady2023advances} and binomial \cite{Marios2016, Hu2019} codes. Recently the original GKP code has received strong renewed attention. Using controls optimized via reinforcement learning (RL), Sivak et al. \cite{Sivak2023} have demonstrated GKP error correction to go beyond the so-called break-even point of the best passive encoding in the same system, extending the lifetime by a factor of 2. This represents by far the longest qubit lifetime extension reached by any platform \cite{Hu2019, CampagneIbarcq2020, Gertler2021, Krinner2022, PhysRevLett.129.030501, Sundaresan2023, Google2023}. However, even though current approaches to QEC for this system are based on protocols that use feedback, the feedback only affects a few operations, and the response depends only on the latest measurement outcome. First attempts to use memory in a handcrafted, constrained way for GKP QEC have shown promising results \cite{Kwok2020}. This raises the question how one would go about discovering QEC strategies that are able to fully employ memory and whether these can significantly boost the performance.\\
\indent Machine learning has proven to be a powerful tool for improving quantum technologies \cite{August2017memory, Porotti2022, Krenn2023, Gebhart2023, Reuer2023}, especially for discovering new strategies for quantum error correction \cite{Torlai2017, Krastanov2017, Foesel2018, Varsamopoulos_2018, Baireuther2018, Liu2019, Baireuther_2019}.  
Recently, Porotti et al.~\cite{Porotti2023} have developed a method for using gradient-ascent pulse engineering (GRAPE) with feedback, known as \textit{Feedback-GRAPE}, that allows for a feedback-based optimal control with a model-based approach. This has been demonstrated to outperform state-of-the-art model-free RL methods in tasks like state purification, state preparation and stabilization, and bosonic code preparation. \\
\indent In this work, we propose an advanced GKP QEC scheme making use of a recurrent neural network (RNN), which is able to provide an elaborate QEC strategy based on the memory of the entire sequence of measurement outcomes. In particular, we use \textit{Feedback-GRAPE} to train the RNN, which suggests an optimal real-time feedback response, adaptively adjusting the parameters of the GKP QEC circuit  (Fig.~\ref{scheme}). Thanks to the memory of the neural network, the discovered real-time strategy is non-Markovian and is able to boost the theoretical performance via non-Markovian feedback (NMF), increasing the logical qubit's lifetime significantly beyond the already powerful standard strategies, by about 100\%. \\

\textit{Theoretical background.} In order to be able to counteract the effect of noise in quantum devices, and thus to correct for errors which would lead to the loss of stored quantum information, different QEC protocols have been developed for the GKP code in the past decades~\cite{Puri2021}. We can conveniently distinguish two classes of QEC strategies: namely, autonomous \cite{PhysRevLett.125.260509} and feedback-based \cite{CampagneIbarcq2020, Sivak2023}. While the former does not include any measurement, the latter relies on a feedback loop based on projective measurements of an ancillary qubit to perform operations dependent on the measurement outcomes. Although both approaches have advantages and disadvantages, feedback-based protocols have shown to have a better performance in experiments than autonomous QEC schemes \cite{CampagneIbarcq2020, Sivak2023}, naturally allowing for further extensions and developments. In our work we focus entirely on the up-to-date best optimized measurement-based protocol: namely, the so-called small-BIG-small (sBs) protocol \cite{PhysRevLett.125.260509,deNeeve2022,Sivak2023}. In particular, we adopt the implementation used by Sivak et al. \cite{Sivak2023}, since it is experimentally feasible and it has been demonstrated to be more versatile for further optimization. This protocol is based on four elementary gates which allow for a universal control of the cavity-qubit system \cite{Eickbusch2022}: namely, the \textit{echoed conditional displacement} $\hat{ECD}_{qc}$ (with complex parameter $\beta$), acting as an entangling gate on both the ancillary qubit and the cavity; the \textit{qubit rotation} $\hat{R}_q$ (with real parameters $\phi, \theta$); the cavity \textit{displacement} $\hat{D}_c$ (with real parameter $\alpha$); and the cavity \textit{virtual rotation} $\hat{VR}_c$ (with real parameter $\theta_{VR}$) \cite{supplement}. Quantum error correction is performed by applying an even number of half-cycles, each of which includes the same fixed sequence of gates (see Fig.~\ref{scheme_QEC}). Initially, the ancilla qubit is prepared in its ground state $| g \rangle $, and then 4 layers of the QEC half-cycle consisting of qubit rotation and cavity displacement are applied. At this time, a measurement on the ancillary qubit is performed, projecting its state along the $z$ axis, returning $g$ ($e$) if projected onto its ground (excited) state $| g \rangle$ ($| e \rangle$).
In our work we neglect leakage errors which could get the qubit to be in higher excited states. After the measurement, the ancilla qubit is reset to its ground state just before the next stage of control sequence takes place. \\
\begin{figure}[h!]
\centering
\includegraphics[width=0.9\linewidth]{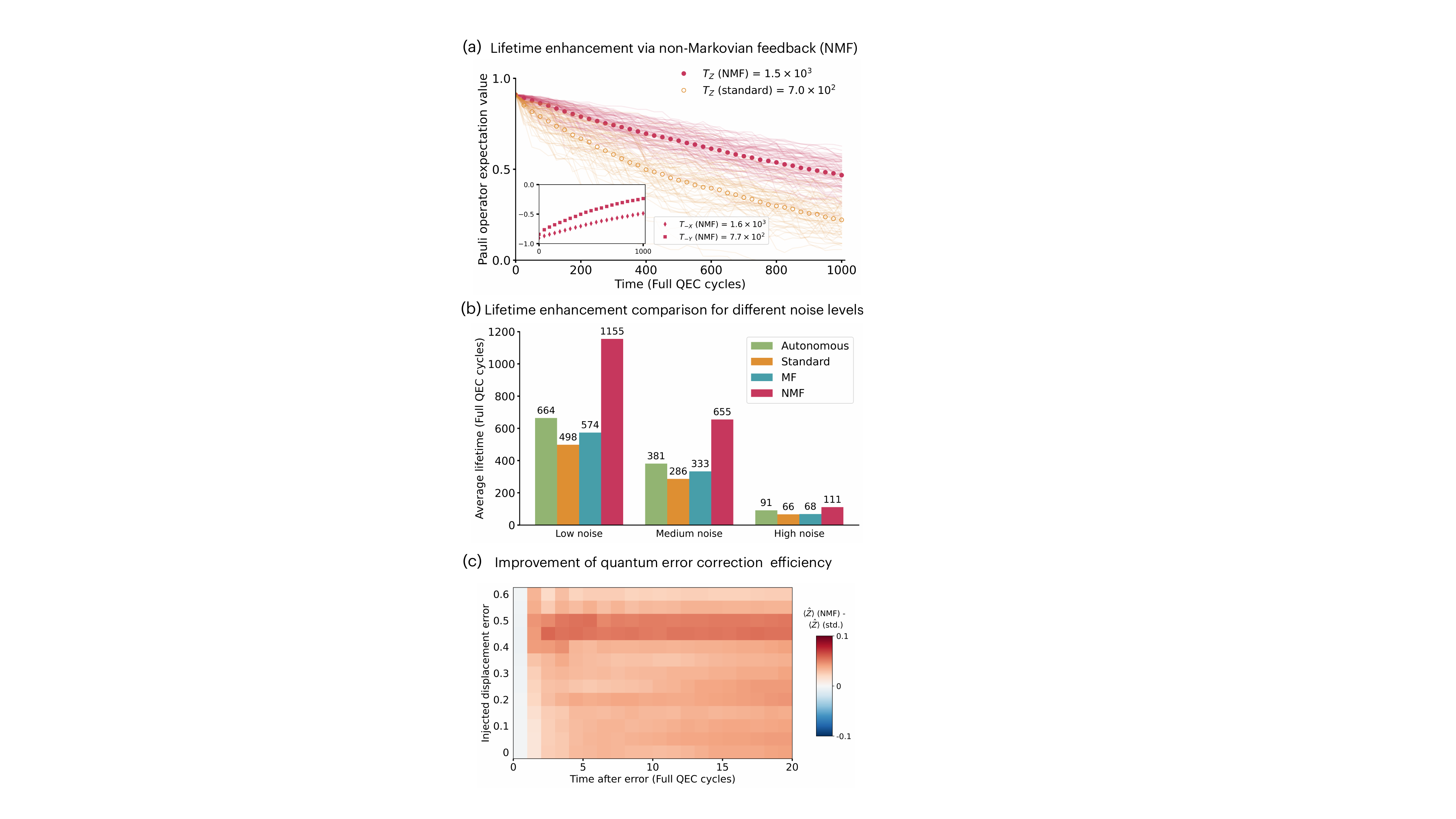}
\caption{Logical qubit's lifetime extension. (a) Comparison of the logical state $\lvert +Z_L\rangle$ time evolution (dots: average over multiple trajectories, continuous faded lines: sampled trajectories averaged over the batch-size) for the standard sBs protocol and the NMF approach advocated in this work. The inset shows the results obtained with NMF for $\lvert - Y_L \rangle$ and $\lvert - X_L \rangle$. (b) Lifetime of average channel fidelity for four implementations of the sBs protocol: autonomous, standard, MF and NMF. The results are obtained for fixed time intervals of each cycle (Fig.~\ref{scheme_QEC}) and different noise levels (characterized by decoherence times $T_1, T_2, T_s$) \cite{supplement}. (c) QEC after injection of a displacement error: difference between the NMF strategy and the standard sBs protocol, after averaging over 1032 trajectories.} \label{results}
\end{figure} 
\indent The QEC task presented can be described as a \textit{quantum observable Markov decision process} (a physical application of \textit{hidden Markov models} and \textit{partially observable Markov decision processes}) \cite{HMM1989, Barry2014, Sivak2022, markov2023}, since it is not possible to (experimentally) access the full quantum state during the process: the only available information is the outcome of the projective measurements of the ancillary qubit entangled with the oscillator. However, each observation alone is insufficient to determine the best parameter set for the applied gates in the following QEC cycle. Therefore, we adopt a RNN to integrate information over time, effectively building an internal representation (or belief state) that approximates the full state of the system. In fact, this capability of RNNs has been often exploited in different complex tasks like natural language processing, time series prediction, and many others with applications to  scientific problems \cite{RICOMARTNEZ1992, GICQUEL19988, 4531750, cho-etal-2014-learning, NIPS2015_07563a3f, Tsai2020}. As a result of the solution adopted, the QEC strategy proposed becomes dependent also on all the past observations: in this sense the strategy is non-Markovian. \\
\indent We implement a realistic simulation of the collective entangled dynamics of the cavity and the ancilla qubit, with fixed time dynamics (Fig.~\ref{scheme_QEC}) and full cycle duration $\tau_{cycle}$. This allows us to compute with high fidelity the probability that a sequence of measurement outcomes is obtained, which is used to train the neural network resulting in a model-aware agent.
We call $m_i$ (with $i \in \mathrm{N}$) the binary measurement outcome on the ancilla qubit entangled with the cavity in the half-cycle $i$, so that $m_i \in \{g, e\}$.
The full sequence of measurement outcomes is contained in the vector $\mathbf{m} = (m_0, m_1 , \dots, m_N)$, and  $P(\mathbf{m})$ is the probability of the full trajectory to occur. 
%The goal of the optimal control algorithm
In order to extend the logical qubit's lifetime, our goal is to maximize a return function $\mathcal{R}$ that we choose to be the fidelity of the final density matrix with respect to the initial state: namely, $\mathcal{R} = \mathcal{F}(\rho_{Z_L}, \rho (T))$, with $\mathcal{F} (\sigma, \rho) = [ \mathrm{Tr} \lbrace (\sqrt{\hat{\sigma}} \hat{\rho} \sqrt{\hat{\sigma}} )^{1/2} \rbrace ]^2$.
Since we deal with probabilistic sequence of outcomes, we have to maximize the weighted-average cumulative return $\langle \mathcal{R} (\mathbf{m}) \rangle_m = \sum_{\mathbf{m}} P(\mathbf{m}) \  \mathcal{R} (\mathbf{m})$ over all the possible trajectories, according to their corresponding probability given.
This is performed using automatic differentiation of the expression
\begin{align}
    \dfrac{\partial \langle \mathcal{R} (\mathbf{m}) \rangle_\mathbf{m} }{\partial \theta} = \Big\langle \dfrac{\partial \mathcal{R} (\mathbf{m}) }{\partial \theta}  + \mathcal{R} (\mathbf{m}) \dfrac{\partial \ln P_\theta(\mathbf{m}) }{\partial \theta} \Big\rangle_\mathbf{m} \ , \label{fGalg}
\end{align}
in order to compute gradients through the multi-step dynamics. 
While the first term on the \textit{rhs} of Eq.~(\ref{fGalg}) is straightforward, the second one is essential to include the dependence of the probability of each trajectory on the parameters vector $\theta$ of the RNN, as described in Ref.~\cite{Porotti2023}. 
The solution provides us with the best weights and biases for the RNN resulting in the optimal feedback-based QEC strategy.\\

\textit{Results.} 
In our numerical simulations \cite{Puviani2024GitHub} we include the dissipative dynamics via the solution of the Lindblad master equation to simulate an approximation of realistic experimental dynamics, using experimental parameters from Ref.~\cite{Sivak2023}, including the qubit measurement and reset time as well as the delay for the application of each gate \cite{supplement}. The training is performed on a fixed initial logical state ($\lvert Z_L \rangle$) with trajectories of 10 full QEC cycles, in the presence of high noise level \cite{supplement}.
The evaluation of the strategy has been performed on 1000 full QEC cycles, thus extending the time to a much longer range than the one used for the training of the neural network agent itself, and on different noise levels. \\
\indent In Fig.~\ref{results}(a), we show the results of QEC with low noise level (with values from the experiment in Ref.~\cite{Sivak2023}) averaged over more than 500 total trajectories for a logical qubit $\lvert +Z_L \rangle$, both with the standard sBs protocol \cite{Sivak2023} (see \cite{supplement}) and with our non-Markovian-feedback approach. The lifetime more than doubles with the use of NMF: in fact it increases from $T_Z (\text{std.})/ \tau_{cycle} = 7.0 \times 10^2$  to $T_Z (\text{NFM})/ \tau_{cycle} = 1.5 \times 10^3$. The full results for all the eigenstates of the Pauli operators and additional plots are present in the Supplemental Material \cite{supplement}.
Moreover, the learned optimal strategy also generalizes to any initial logical qubit state without re-training, maximizing each corresponding Pauli operator expectation value and extending the qubit's lifetime. Indeed, a comparable result is obtained (inset of Fig.~\ref{results}(a)) using the same neural network with the initial logical state $\lvert - X_L \rangle$, while for the $\lvert - Y_L \rangle$ we get a lower value of $T_{-Y} (\text{NMF})/ \tau_{cycle} = 7.7 \times 10^2$, which is an expected feature of the square GKP code \cite{CampagneIbarcq2020, Sivak2023}. Moreover, we show that a non-Markovian-feedback strategy based on a RNN is necessary to boost the performance of QEC. In fact, an instantaneous Markovian feedback (MF) implemented with a feed-forward neural network, which is capable of providing distinct parameters' values according to the latest measurement output only, performs comparably with the standard approach with no feedback (Fig.~\ref{results}(b)). These results hold even for very different noise levels: from the low noise in Ref.~\cite{Sivak2023} to the higher noise of the experiment in Ref.~\cite{CampagneIbarcq2020}, as well as in the presence of imperfect gates \cite{supplement}. \\
\indent In addition, we notice that our NMF approach considerably outperforms other approaches, including the autonomous QEC scheme which has a shorter protocol duration \cite{AutonomousQEC2024, supplement}. Apparently, measurement based protocols are not intrinsically outperforming the autonomous one under all circumstances, but they become strongly advantageous when memory is exploited, as in our scheme. 
We also compare the standard and the NMF sBs protocols in their ability to correct deliberately injected errors (Fig.~\ref{results}(c)): by injecting displacement errors to the initial $| + Z_L \rangle$ state before QEC, we compare the ability of the two methods to recover the initial logical information, namely the expectation value of the logical Pauli operator $\hat{Z}_L$. The plot in Fig.~\ref{results}(c) shows that the performance difference of our approach and the standard sBs is always positive, confirming that our approach allows to reach higher values in a shorter time compared to the standard QEC strategy. \\
\indent We now analyze the strategy discovered by the RNN with non-Markovian feedback,
\begin{figure}[h!]
\centering
\includegraphics[width=1.0\linewidth]{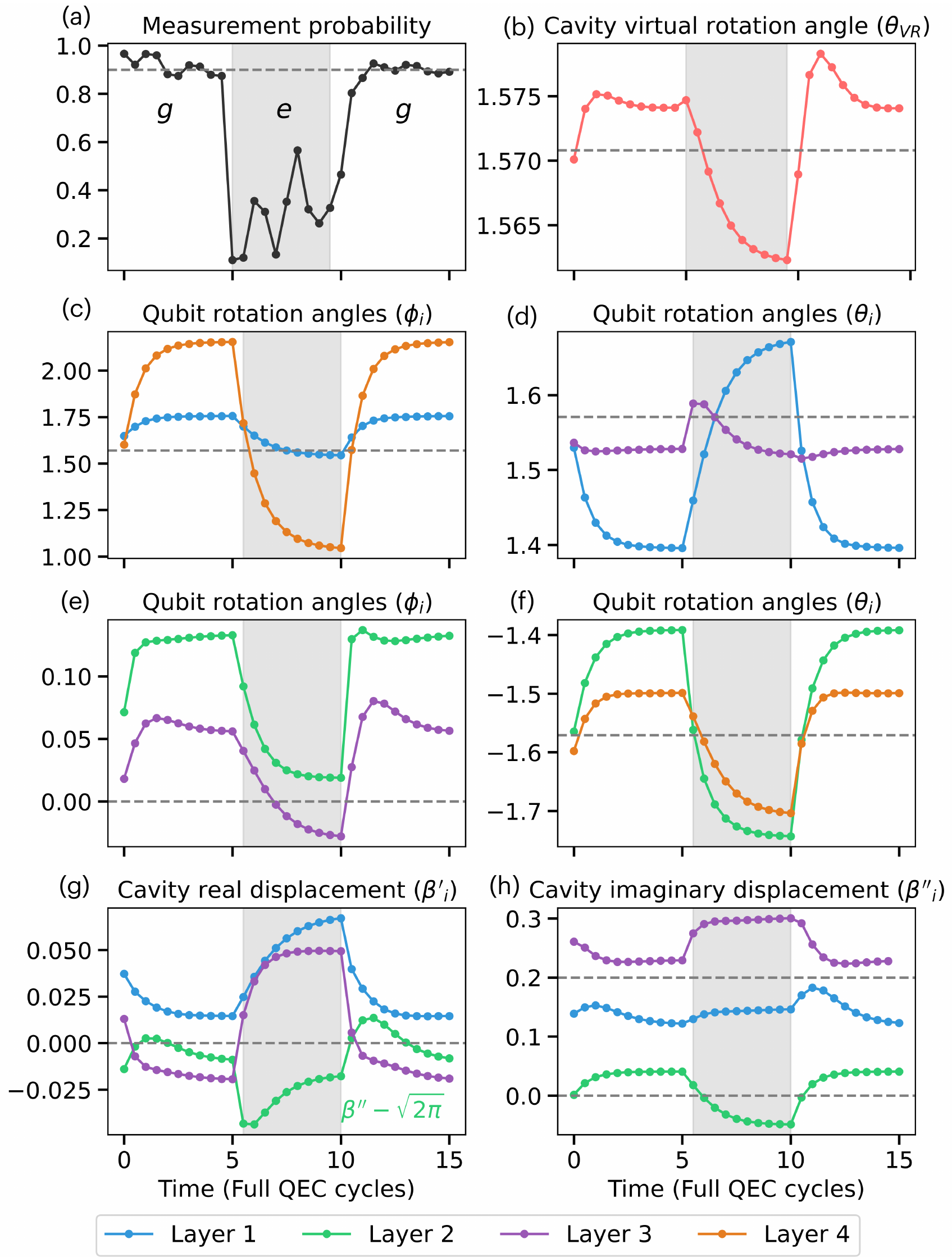}
\caption{Optimized non-Markovian parameters. (a) Measurement probabilities for a given QEC trajectory, with fixed measurement outcomes. The value $g$ ($e$) corresponds to an outcome indicating the ancilla qubit state in $| g \rangle$ ($| e \rangle$). The dashed line indicates the probability $0.9$. (b) Virtual rotation angle applied at the end of each QEC cycle. The vertical shaded area represents the cycle's timestep at which the measurement $m = e$ is obtained and the virtual rotation is applied. (c)-(h) Parameters entering the 4 layers of the QEC cycle in Fig.~\ref{scheme_QEC} suggested by the RNN. The vertical shaded areas represent the cycles in which the RNN is given the $e$ measurement as input. The dashed horizontal lines are a guide for the eyes indicating the corresponding constant values of the standard sBs protocol.} \label{strategy}
\end{figure}
showing how memory is exploited to outperform QEC. In Fig. \ref{strategy} we present one case of a trajectory for the initial logical state $| + Z_L \rangle$ with noisy dynamics and fixed post-selected measurement outcomes that we prescribe (sequence of outcomes: 10 $g$, 10 $e$ and 10 $g$). This allows us to track the values of the 15 parameters optimized with the NMF approach in an easily interpretable situation of subsequent $g$ and $e$ measurement outcomes. 
For most of the parameters we can notice that the initial values are relatively close to the standard one (represented by dashed lines) \cite{Sivak2023}, and then they evolve over time with an alternating exponential-like behavior according to the previous measurement outcomes. As a result, it would be possible to convert this strategy into an analytical expression, which would be easier to implement in an actual experimental realization \cite{supplement}.
\\
\indent We also find that our optimal strategy has the same probability of the qubit to be found in the ground state $g$ after the measurement, $p(g) \approx 0.9$, as in the standard approach (see Fig.~\ref{strategy}(a)). This observation is important because as explained in Ref.~\cite{Sivak2023} this is experimentally desirable when implementing the sBs measurement-based protocol. Other strategies that we found, instead, occasionally exhibit a longer lifetime in some conditions, but they are less generalizable to different noise levels and show a low probability $p(g)$ averaged over time, thus becoming less auspicable for an experimental implementation. 
We present more systematic results for other agents, the Markovian-feedback strategy, as well as for biased-noise gates, different dynamics and in the presence of cavity dephasing in the Supplemental Material \cite{supplement}. \\

\textit{Conclusion.} 
In this work, we have proposed a measurement-based QEC scheme for GKP codes based on non-Markovian feedback. In particular, we have used a recurrent neural network in order to optimize all the QEC operations based on the full history of previous measurement outcomes. We have also shown that our QEC scheme remarkably outperforms current autonomous (without feedback) and measurement-based (with Markovian feedback) QEC strategies in different conditions of noise level and dynamics. To reach this result we have used the recently proposed Feedback-GRAPE  \cite{Porotti2023} (Gradient Ascent Pulse Engineering with Feedback) method to train the RNN exploiting a model-based approach. \\
\indent More generally, our work demonstrates the power of optimizing feedback protocols in the shape of variational quantum circuits \cite{Barkoutsos2020, PhysRevResearch.3.023092, Cerezo2021} with in-sequence discrete measurements.
We additionally envision the possibility to extend this non-Markovian approach to any measurement-based QEC strategy for other bosonic codes, with the model-based approach enabled by the Feedback-GRAPE algorithm.  \\

\textit{Acknowledgments}. Fruitful discussions with V. Peano and V. V. Sivak are thankfully acknowledged. The research is part of the Munich Quantum Valley, which is supported by the Bavarian state government with funds from the Hightech Agenda Bayern Plus.\\

\end{document}

\title{Supplemental Material for \\ "Non-Markovian feedback for optimized quantum error correction"}

\author{Matteo Puviani}
\email{matteo.puviani@mpl.mpg.de}
\affiliation{Max Planck Institute for the Science of Light, 91058 Erlangen, Germany}

\author{Sangkha Borah}
% \email{}
\affiliation{Max Planck Institute for the Science of Light, 91058 Erlangen, Germany}
\affiliation{Department of Physics, Friedrich-Alexander Universität Erlangen-Nürnberg, 91058 Erlangen, Germany}

\author{Remmy Zen}
% \email{}
\affiliation{Max Planck Institute for the Science of Light, 91058 Erlangen, Germany}

\author{Jan Olle}
% \email{}
\affiliation{Max Planck Institute for the Science of Light, 91058 Erlangen, Germany}

\author{Florian Marquardt}
\email{florian.marquardt@mpl.mpg.de}
\affiliation{Max Planck Institute for the Science of Light, 91058 Erlangen, Germany}
\affiliation{Department of Physics, Friedrich-Alexander Universität Erlangen-Nürnberg, 91058 Erlangen, Germany}

\maketitle

\section{The GKP code} 
In contrast to multi-qubit encoding, where the logical qubit information is stored into many physical qubits allowing to implement fault-tolerant operations and to correct for quantum decoherence errors, bosonic codes exploit the theoretically infinite Hilbert space of Fock states of a bosonic system. 
There are different bosonic codes which we can consider, including the Pegg-Barnett \cite{Barnett1986}, cat (kitten) \cite{Cochrane1999}, Gottesman-Kitaev-Preskill (GKP) \cite{GKP2001, PhysRevLett.125.260509, CampagneIbarcq2020, Puri2021, brady2023advances} and binomial \cite{Marios2016, Hu2019} code.
In this work, we consider a single-mode GKP state encoding a single logical qubit. 

In the superconducting circuit platform, these codes can be physically realized using an optical superconducting cavity, coupled to an ancilla transmon qubit and a readout resonator. The ancillary qubit is entangled with the cavity state and is used to control the bosonic code itself. In the case of measurement-based quantum error correction (QEC) protocols, projective measurements are performed on the qubit in order to extract weakly information from the cavity, without destroying the encoded logical state.

\begin{figure}[hb!]
\centering
\includegraphics[width=0.5\linewidth]{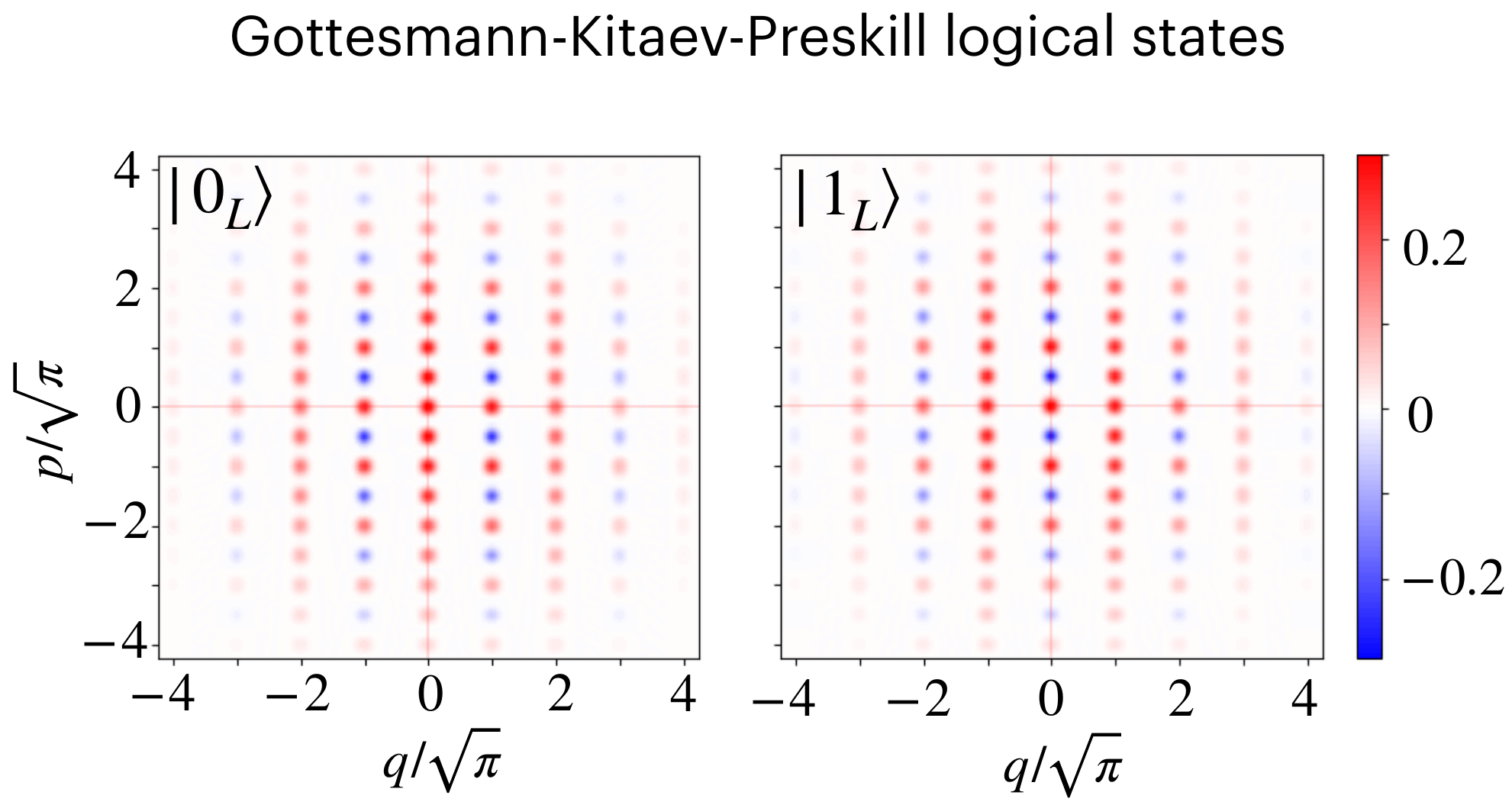}
\caption{Wigner functions of the GKP physical code for the logical qubits $0_L$ and $1_L$, respectively. The two logical grid states differ only by a displacement of $\sqrt{\pi}$ along $q$.}
\label{scheme_GKP}
\end{figure}

\subsection{Pauli operators and stabilizers} \label{logical_ops}
Here we describe the theoretical framework to define and characterize the GKP codes, focusing on the case of a cavity with one single oscillator mode.
We start defining the cavity displacement operator 
\begin{align}
    \hat{D} (\zeta) = e^{\zeta \hat{a}^\dagger - \zeta^* \hat{a} },
\end{align}
where $\hat{a}(\hat{a}^{\dagger})$ is the bosonic annihilation (creation) operator. \\
There are two stabilizers, $\hat{S}_X = \hat{X}_L^2 = \hat{D}(2 \alpha)$ and $\hat{S}_Z = \hat{Z}_L^2 = \hat{D}(2 \beta)$, which define the entire GKP codespace spanned by the two logical states $|0_L \rangle$ and $| 1_L \rangle$, i.e.: $\mathcal{C} = \big\langle | 0_L \rangle, | 1_L \rangle \big\rangle$. Every state in $\mathcal{C}$ can be written as a linear combination of this two logical states (which form a basis) and is a simultaneous eigenstate of both the stabilizers with eigenvalue $+1$, as the stabilizers also commute with each other. This physically means that every logical qubit is characterized by the periodicity $2 \alpha$ and $2 \beta$ in the $\lbrace q,p \rbrace$ phase-space. \\
The logical Pauli operators are instead defined as $\hat{X}_L = \hat{D}(\alpha)$ and $\hat{Z}_L = \hat{D}(\beta)$, such that their eigenstates are the logical states $\pm X$ ($| \pm X_L \rangle$) and $\pm Z$ ($| \pm Z_L \rangle$), respectively. The operator $\hat{Z}_L$ is used to define the basis logical states: $| + Z_L \rangle = | 0_L \rangle$ and  $| - Z_L \rangle = | 1_L \rangle$. Physically, the eigenstates of the logical Pauli operators with opposite eigenvalues are transformed into each other with the application of a displacement $\hat{D} (\alpha)$ or $\hat{D}{\beta}$, respectively. For this reason, the Wigner function of a logical state in this encoding is characterized by dots forming a periodic lattice, which is why the GKP code is also referred to as the \textit{grid code} [Fig. \ref{scheme_GKP}]. \\

\subsection{Logical qubit states}
The values of the complex code parameters $\alpha$ and $\beta$, which enter the definition of the Pauli operators and stabilizers, are not unique, but they are constrained by a relation according to anti-/commutation relation of the logical operators: 
\begin{align}
    \alpha^* \beta - \alpha \beta^* = i \pi \ , 
\end{align}
or 
\begin{align}
    \alpha^* \beta - \alpha \beta^* = 2 i \pi \ , 
\end{align}
respectively. We choose here the condition for which the logical operators  anti-commute. We can now define the logical qubits encoding $| 0_L \rangle = | + Z_L \rangle$ and $| 1_L \rangle = | - Z_L \rangle$ as \cite{Puri2021}
\begin{align}
    | 0_L \rangle &= \sum_j | 2 j \sqrt{\pi} \rangle_Q \ , \\
    | 1_L \rangle &= \sum_j | (2 j + 1) \sqrt{\pi} \rangle_Q \ ,
\end{align}
with the eigenstates of the position operator $\hat{Q} = i (\beta^* \hat{a} - \beta \hat{a}^\dagger)/ \sqrt{\pi}$, or equivalently 
\begin{align}
    | 0_L \rangle &= \mathcal{N} \sum_{k,l} e^{- i \pi k l} | 2 k \alpha + l \beta \rangle \ , \label{logical0} \\ 
    | 1_L \rangle &= \mathcal{N} \sum_{k,l} e^{- i \pi (k l + l/2)} | (2 k +1) \alpha + l \beta \rangle \ , \label{logical1}
\end{align}
using a linear combination of the coherent states of the cavity oscillator, $\mathcal{N}$ being a normalization coefficient. \\
Since the anti-commutation relation is the only constrain over the two parameters of the code, it is possible to define GKP codes that have different symmetries in the phase space according to the values of $\alpha$ and $\beta$, such as the square, rectangular and hexagonal code. 
For the square code, characterized by a square lattice periodicity of the Wigner function, we have $\alpha= \sqrt{\pi/2}$ and $\beta = i \alpha$. Therefore $\alpha$ is the unitary lattice displacement along the $\hat{q}$ direction, while $\beta$ is the corresponding one along $\hat{p}$. The rectangular code is generalized as $\alpha = l \sqrt{\pi /2}$, $\beta = i \sqrt{\pi / 2} / l$, where $l$ is a free scaling parameter, while the hexagonal code has $\alpha = \sqrt{\pi / \sqrt{3}}$ and $\beta = e^{2 i \pi /3}\sqrt{\pi / \sqrt{3}}$. \\
In this work, we have focused entirely on the square GKP code.

\subsection{Finite-energy GKP code}
The ideal GKP logical qubit encoding as described in the previous section, however, is not physical since it requires an infinite combination of coherent states and therefore an infinite amount of photons and energy to be implemented. Thus, a real finite-energy GKP code can be described by truncating the number of coherent states up to a maximum value $M$: physically this corresponds to considering a finite-size grid in phase space instead of an infinite periodic lattice, as represented in Fig. \ref{scheme_GKP}. Mathematically, this can be described by a convolution of a gaussian envelope with the ideal logical state in Eqs. (\ref{logical0}) and (\ref{logical1}):
\begin{align}
    | \tilde{\mu}_L \rangle \propto \ e^{- \Delta^2 \hat{a}^\dagger \hat{a}} | \mu_L \rangle \ , 
\end{align}
with $ \mu = 0,1$ for the corresponding logical qubit states. The ideal limit is obtained for $\Delta \rightarrow 0$. \\
As we can notice, the gaussian envelope introduces a cut-off both in real and phase space, as well as in the Hilbert space of Fock states, providing an upper limit to the actual number of photons involved. Therefore, it is possible to accurately simulate the dynamics of a finite-energy GKP code imposing a large but finite number of Fock states to be considered. In our work we adopted an upper limit of the Fock space of $N_{max} = 100$. In our simulations we used the value for the initial logical state $\Delta = 0.34$, corresponding to an average photon number $\langle \hat{n} \rangle \approx 5$, consistently with recent experiments \cite{Sivak2023}. \\
However, the main consequence of this is that the GKP code is no longer invariant under discrete translations, but only approximately and within a few translations: therefore, one additional challenge of QEC is to keep the grid centered in phase space around the $(q,p) = (0,0)$ point.\\
It is also possible to introduce \textit{finite-energy} or \textit{approximate} Pauli operators and stabilizers, in order to normalize them to the GKP finite energy \cite{Puri2021}: however, we did not use such operators, but rather the standard definition provided in the previous section.

\section{GKP Quantum error correction}

\subsection{Dynamics, noise and errors of the GKP code} \label{noise}

In order to be able to counteract the effect of photons' dissipation in the quantum cavity by acting and applying controls on it, the cavity itself is coupled to a two-level system that acts as control qubit. However, the qubit that is adopted as ancilla is also affected by quantum noise, namely relaxation, dephasing and leakage events. The former two are responsible for the decoherence of the qubit, while the latter is the most harmful error and is due to the excitation to higher levels in an actual multi-level system. A protocol must be developed to correct for these additional qubit errors that would affect the cavity after entanglement, as these errors (especially the leakage error) can be the most detrimental to GKP coding. A readout resonator completes the hardware setup that enables the measurement of the ancilla qubit: a strong measurement of the qubit in this configuration is reflected into a weak measurement of the cavity so that the logical quantum information contained therein is not destroyed.  
In our treatment we will not consider the qubit leakage and the associated errors, as well as SPAM errors: this would have equal impact on all the measurement-based protocols and would not significantly affect the results from a quantitative point of view. In fact, in an experimental setup such as the one in Ref.~\cite{Sivak2023}, the probability of a leakage error is $p_{leak.} < 0.01$, and it can be reset in the same way as the excited qubit state.

We consider the ground state of the ancillary qubit $ | g \rangle = \begin{pmatrix} 
1 \\ 
0 
\end{pmatrix}$, so that the Pauli operators $\hat{\sigma}_- = \left( \hat{\sigma}_x - i \hat{\sigma}_y \right)/2$ and $\hat{\sigma}_+ = \left( \hat{\sigma}_x + i \hat{\sigma}_y \right)/2$ act as creation and annihilation ladder operators, respectively. We define the dissipator of a damping operator $\hat{A}$ applied to the density matrix $\hat{\rho}$ as
\begin{align}
    \mathcal{D} [\hat{A}] \hat{\rho} = \hat{A} \hat{\rho} \hat{A}^\dagger - \left( \hat{A}^\dagger \hat{A} \hat{\rho} + \hat{\rho} \hat{A}^\dagger \hat{A} \right) /2 \ , \label{Lindblad}
\end{align}
in order to be able to write the time evolution of the density matrix with the Lindblad master equation for the cavity-ancilla system in the following form:
\begin{align}
    \dot{\hat{\rho}} = - i \left[ \hat{H}, \hat{\rho} \right] + \dfrac{1}{T_s} \mathcal{D}[\hat{a}] \hat{\rho}  + \dfrac{2}{T^{\text{white}}_{\phi,c}} \mathcal{D}[\hat{a}^\dagger \hat{a}] \hat{\rho} +  \dfrac{1}{T_1} \mathcal{D}[\hat{\sigma}_+] \hat{\rho} + \dfrac{2}{ T_\phi}\mathcal{D}[\hat{\sigma}_z/2 ] \hat{\rho} \ ,
\end{align}
where $T_s$ is the cavity single-photon lifetime, $T_1$ is the qubit relaxation time, $T_2$ the qubit full dephasing time, while $T_\phi = \dfrac{1}{1/ T_2 - 1/2 T_1}$ is the pure dephasing time of the qubit. Similarly, $T^{\text{white}}_{\phi,c} = \dfrac{1}{1/ T_{2,c} - 1/2 T_s}$ is the intrinsic white-noise pure dephasing time of the cavity, $T_{2,c}$ being the cavity full dephasing time. \\
The intrinsic pure dephasing (here named \textit{white noise dephasing}) time of stand-alone cavities not coupled to qubits (coming from frequency fluctuations) typically ranges around $T_{\phi,c}^{\text{white}} = 110-150$ ms \cite{Eickbusch2022, AutonomousQEC2024, goldblatt2024}, which makes this source of errors negligible at the scale of the GKP code lifetime \cite{Sivak2023}. For this reason, in our simulations we did not consider any pure dephasing of the cavity in addition to the part that results from the cavity decay. However, in the community it is indeed known that additional pure dephasing sourced are present in the experimental realizations: We discuss this topic more in detail in section \ref{Dephasing} of this Supplemental Material. \\
The Hamiltonian of the system is $\hat{H}$, and it includes a coupling and a Kerr term:
\begin{align}
    \hat{H} = \dfrac{1}{2} \chi \hat{a}^\dagger \hat{a} \hat{\sigma}_z + \dfrac{1}{2} K \left(\hat{a}^\dagger \hat{a} \right)^2 \ , 
\end{align}
where $\chi$ is the first order dispersive shift term and $K$ is the Kerr constant. Since the dynamics generated by the Hamiltonian is used to model the pulses to give rise to the gates of the QEC circuit \cite{CampagneIbarcq2020}, in our work we neglected its contribution to the dynamics in Eq.(\ref{Lindblad}) because we adopted perfect instantaneous parameterized gates followed by idling dissipative dynamics, without modelling the electromagnetic pulses.

For the purpose of quantum error correction, we can add projective quantum measurements on the ancillary qubit, which act as weak (syndrome) measurement on the cavity state providing information on whether an error has occurred \cite{Sivak2023}. Since the quantum measurements have an intrinsically statistical nature, we stochastically sample the measurement's outcome with a Monte-Carlo algorithm. The density matrix of the full quantum system (cavity and ancilla) after a measurement is 
\begin{align}
    \hat{\rho} (t_+) = \sum_{m = \pm 1} r_m \ \hat{P}_m \ \hat{\rho} (t_-) \ \hat{P}^\dagger_m \ , \label{meas_eq}
\end{align}
with the sampling algorithm
\begin{align}
    r_m = 
    \begin{cases}
    1, & \text{if $p_m > s$}\\
    0, & \text{otherwise}
     \end{cases}
\end{align}
where $s \in [0,1]$ is sampled from a uniform distribution, $p_m = \mathrm{Tr} \lbrace \hat{P}_m \hat{\rho} \hat{P}^\dagger_m \rbrace$ is the probability of measurement $m \in \{g,e\}$, and $\hat{P}_m = \hat{\mathbb{1}}_c \otimes | m \rangle_q \langle m |$ is the projection operator onto the $| m \rangle$ qubit state. Here we introduced the subscripts $q$ and $c$ to indicate the qubit and the cavity, respectively.\\
The Lindblad master equation (\ref{Lindblad}) including the discrete stochastic measurements (\ref{meas_eq}) becomes overall probabilistic. Moreover, we notice that both of them define the evolution of the density matrix immediately after the measurement $\hat{\rho} (t_+)$ in a Markovian way, as they only depend on the density matrix $\hat{\rho}(t_-)$ at the step immediately before.

\subsection{Physical gates}
To implement the small-BIG-small QEC protocol as represented in Fig.~\ref{scheme_QEC} in the main text following Ref.~\cite{Sivak2023}, we adopted a set of universal physical gates acting on the cavity [the cavity virtual rotation $\hat{VR}_c (\theta_{VR})$ and the cavity displacement $\hat{D}_c (\alpha)$], on the ancillary qubit [the the qubit rotation $\hat{R}_q (\phi, \theta)$], and on both [the echoed conditional displacement $\hat{ECD}_{qc} (\beta)$]. For these gates we adopt the definition given in Ref.~\cite{Eickbusch2022}, with the echoed conditional displacement defined as
\begin{align}
    \hat{ECD}_{qc} (\beta) = \hat{D} (\beta / 2) \otimes \hat{\sigma}_- + \hat{D} (- \beta / 2) \otimes \hat{\sigma}_+ \ ,
\end{align}
where $\beta \in \mathbb{C}$. The operators $\hat{\sigma}_{\pm}$ are acting on the ancillary qubit, while the displacement $\hat{D}$ acting on the cavity is defined as
\begin{align}
    \hat{D}_{c} (\beta) = \exp \lbrace \beta \hat{a}^\dagger - \beta^* \hat{a}   \rbrace \ .
\end{align}
The ancillary qubit rotation is given by
\begin{align}
    \hat{R}_q (\phi, \theta) = \exp \Big\lbrace - i \frac{\theta}{2} ( \hat{\sigma}_x \cos{\phi} + \hat{\sigma}_y \sin{\phi}) \Big\rbrace.
\end{align}
The cavity virtual rotation that allows to rotate the $q$ and $p$ axis in phase space is given by
\begin{align}
    \hat{VR}_c (\theta_{VR}) = \exp \lbrace i \theta_{VR} \hat{a}^\dagger \hat{a} \rbrace \ . 
\end{align}
This is used in order to keep the same fixed parameters in the QEC circuit without swapping between real and complex parameters' values.

\subsection{small-BIG-small QEC protocol} \label{sBsprotocol}
\begin{figure}[h!]
\centering
\includegraphics[width=1.0\linewidth]{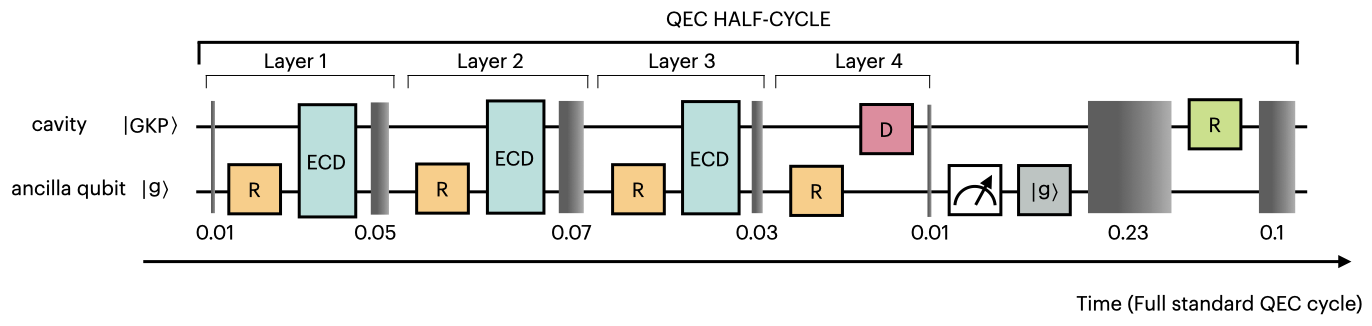}
\caption{Scheme showing the half-cycle measurement-based sBs QEC protocol as adopted in the main paper and shown in Fig.~\ref{scheme_QEC}, with the dynamics represented by the shaded grey areas, with the time indicated in units of the full QEC cycle.} \label{scheme_time}
\end{figure}

In order to perform QEC we adopted the so-called small-BIG-small (sBs) protocol with measurement-based feedback as described in Ref.~\cite{Sivak2023}. In order to simulate the realistic dynamics we modelled the circuit with perfect instantaneous gates followed by idling dissipative dynamics solving the Lindblad master equation (\ref{Lindblad}), neglecting the Hamiltonian's dynamics.
As a consequence, the feedback coming from the measurement, which is used in the experiments to reset the ancilla and to determine the correct angle for the cavity virtual rotation gate, is not neede here in our modeling. 
For the time dynamics during the QEC circuit, we considered a realistic model based on experiment in Ref.~\cite{Sivak2023}: namely, we assumed a total duration $\tau_{\text{cycle}} = 10 \ \mu$s, so that the half-cycle shown in Fig.~\ref{scheme_time} will have a duration of $5 \ \mu$s. The absolute and relative times of each step are shown in Table \ref{table_time}.

The initial condition of the QEC protocol comprises a cavity with a prepared GKP code, and an ancillary transmon qubit set in its ground state, $| g \rangle$. The sBs standard protocol consists of 4 layers of qubit ancilla rotation and echoed conditional displacement, which entangles the cavity and the qubit states (Fig.~\ref{scheme_time}). In the fourth layer the entangling gate is replaced by a simple displacement gate acting on the cavity. Then, a projective measurement (POVM) takes place on the ancillary qubit, which projects its state onto either the ground ($| g \rangle$) or the excited ($| e \rangle$) state. After that, the qubit is reset to its ground state and a virtual rotation is applied on the cavity, in order to switch between the position and momentum space variables. We notice here that we did not include the Hamiltonian dynamics in the time evolution, and the reset is performed numerically and not with gates or pulses. As a consequence, the feedback that is usually applied in experiments for the qubit reset and to compensate the Hamiltonian dynamics in the cavity virtual rotation does not apply to our numerical simulations.

\begin{table}[h!]
\begin{tabular}{l*{3}{ccc}}
\hline
Dynamics  & \ \  Duration time ($\mu$s) \ \ &  \ \ Duration time ($\tau_{cycle}$) \\
\hline
Entering cycle & $0.1 \ \mu$s & 0.01 \\ 
Layer 1 & $0.5 \ \mu$s & 0.05 \\ 
Layer 2 & $0.7 \ \mu$s & 0.07 \\ 
Layer 3 & $0.3 \ \mu$s & 0.03 \\ 
Layer 4 & $0.1 \ \mu$s & 0.01 \\ 
Measurement and reset & $2.3 \ \mu$s & 0.23 \\ 
Virtual rotation and idling & $1 \ \mu$s & 0.1 \\ \hline
\end{tabular}
\caption{Time duration of the dissipative dynamics in half QEC cycle shown in Fig.~\ref{scheme_time}.} \label{table_time}
\end{table}

\section{Feedback-GRAPE optimization with non-Markovian feedback}

\subsection{Feedback-GRAPE} \label{theoryFG}
Feedback-GRAPE is an algorithm recently developed by Porotti et. al~\cite{Porotti2023} for Gradient Ascent Pulse Engineering (GRAPE) with feedback. It works for a quantum system with feedback that is subject to repeated strong projective measurements at times $t_{1,}t_{2,}...,t_{N}$ with the corresponding measurement results $m_{j,}~j\in [1, N].$ The task of Feedback-GRAPE is to find the accurate feedback control functions $F^{j}_{\theta}(\mathbf{m}_j)$ for all possible measurement results $\mathbf{m}_{j}=\{m_{1,}m_{2,}..., m_{j} \}$, as a feedback strategy parameterized with the trainable parameters $\theta$ that are optimized by gradient ascent. It is assumed that $F^{j}_{\theta}$ is differentiable with respect to $\theta,$ which can be provided by a neural network or alternatively a lookup table. In order to optimize them, the task would be to to maximize a cumulative total reward $R$ (this is called return in the nomenclature of reinforcement learning). In a state preparation task, for example, this could be the final fidelity with respect to a target state ${\hat \sigma}$, $\mathcal R(\mathbf{m})=\left( {\rm tr} \sqrt{\sqrt{\hat \sigma} {\hat\rho}(T|\mathbf{m}) \sqrt{\hat \sigma}} \right)^2$, where $\mathbf m$ represents a particular measurement sequence, which would finally be averaged over all possible sequences of measurement results to yield $\bar{\mathcal R}=\left\langle {\mathcal R}(\mathbf{m}) \right\rangle_\mathbf{m}$. For strong projective qubit measurements, as used in the present work, however, the probabilities $P$ for the different measurement results themselves depend on all controls $F^j_{\theta}$ that were applied in the previous time intervals, which must be taken into account when analyzing the gradients with respect to $\theta$. Thus
\begin{equation}
   \langle \mathcal R(\mathbf{m}) \rangle_\mathbf{m} = \sum_\mathbf{m} P(\mathbf{m}) \mathcal R(\mathbf{m}) \,, 
\end{equation}
where $P(\mathbf{m})$ is the probability for measurement outcomes $\mathbf{m}$. Thus, when evaluating $\partial \langle R(\mathbf{m}) \rangle_\mathbf{m} / \partial \theta$, we will get two contributions, 
\begin{equation}
    \frac{\partial [ \mathcal R(\mathbf{m}) P(\mathbf{m})]} {\partial \theta} =  P(\mathbf{m})\frac{ \partial \mathcal R(\mathbf{m})} { \partial \theta} + \mathcal R(\mathbf{m}) \frac{\partial P(m)}{ \partial \theta} \,.
\end{equation}
To enable stochastic sampling of the second term, we rewrite it using 
\begin{equation}
\frac{\partial P(\mathbf{m})}{\partial \theta} = P(\mathbf{m}) \frac{\partial \ln P(\mathbf{m})}{\partial \theta} \,,
\end{equation}
which leads to
\begin{equation}
    \frac{\partial \left\langle \mathcal R(\mathbf{m}) \right\rangle_\mathbf{m}}{\partial \theta} = \left\langle \frac{\partial \mathcal R(\mathbf{m}) } {\partial \theta} \right\rangle_\mathbf{m} + \left\langle \mathcal R(\mathbf{m}) \frac{\partial \ln P_{\theta}(\mathbf{m})}  {\partial \theta} \right\rangle_\mathbf{m} \,,
\end{equation}
where the $P_{\theta}(\mathbf{m})$, with the explicit parameter dependence $\theta$ represents the probability of getting the full sequence of measurement outcomes $\mathbf{m}=(m_1,m_2,\ldots)$. The log-likelihood term can be conveniently computed as the cumulative sum over each trajectory. The  gradients can then be taken for a batch of trajectories.

\subsection{Numerical implementation}
The feedback-GRAPE algorithm relies on a gradient-based optimization: therefore we used \textsc{TensorFlow} \cite{tensorflow2015-whitepaper} to implement automatic differentiation through the whole multi-step dynamics, allowing for an efficient setup which can run in parallel on a GPU to speed up the calculations. 
Moreover, in order to numerically simulate the dissipative dynamics of the system, we developed a custom TensorFlow time-evolution simulator implementing the Runge-Kutta-4 algorithm, which can be run in parallel also on a GPU.

Even though the implementation of a look-up table containing the optimal parameters for a given measurement outcomes sequence would be optimal and easier to implement, this would however be limiting as the number of parameters of the look-up table increase exponentially with the number of QEC cycles, and therefore with time, as shown in \cite{Porotti2023}. Therefore, when evaluating trajectories with many cycles the effective memory of the look-up table and its performance would be limited. 
For this reason, we adopted a recurrent neural network (RNN, whose hyperparameters are shown in Table \ref{table_RNN}) which can be used to suggest parameters even at time scales much longer than the ones used for training. In fact, the RNN responds in a non-Markovian way to the measurement outcomes, building an internal representation (or \textit{belief state}) of the state of the quantum system. However, the update of the RNN's hidden state only depends on the current state and the current measurement outcome, making this update itself Markovian. 
For comparison we also used a feed-forward neural network (FNN), whose hyperparameters are provided in Table \ref{table_FNN}. In this latter case, however, the response is not subject to memory and therefore Markovian, as it will be explained in the results' section.

\begin{table}[hbt]
\begin{tabular}{l*{6}{cccccc}}
\hline
Parameter  & Value \\
\hline
RNN cell & GRU \\
Neurons & [10, 256, 256, 15] \\ 
Input shape & [batch$\_$size,1,1] \\ 
RNN cell activation & tanh \\ 
Dense layers activation & tanh \\ 
Initializer & Random Uniform \\ 
Initial bias & 0.01 \\ \hline
\end{tabular}
\caption{Hyperparameters of the RNN used, with batch$\_$size = 6. The core of the network is the memory provided by the \textit{Gated Recurrent Unit} (GRU)  in the first layer of the neural network.}  \label{table_RNN}
\end{table}

\begin{table}[hbt]
\begin{tabular}{l*{6}{cccccc}}
\hline
Parameter  & Value \\
\hline
Neurons & [256, 256, 15] \\ 
Input shape & [batch$\_$size,1,1] \\ 
Dense layers activation & tanh \\ 
Initializer & Random Uniform \\ 
Initial bias & 0.01 \\ \hline
\end{tabular}
\caption{Hyperparameters of the FNN used, with batch$\_$size = 8.}  \label{table_FNN}
\end{table}

The training of the RNN to suggest the gates parameters from scratch was not successful: therefore we provided as initial values the optimal parameters $\theta_i$ used in Ref.~\cite{Sivak2023} (Table \ref{table_sBs}), allowing the neural network to provide at each step corrections $\delta \theta_i$, applying to the gates the final values $\theta_i + \delta \theta_i$. The gates' parameters which are suggested are in total 15 for each cycle: the rotation angles $\phi$ and $\theta$ of the qubit rotations of each layer, the real and imaginary part of the $ECD$, $\beta'$ and $\beta''$, respectively, in layers 1-3, and the virtual rotation angle $\theta_{VR}$. The range allowed for the corrections $\delta \theta_i$ is $[-2,2]$, except for the virtual rotation for which $\delta \theta_{VR} \in [-1,1]$. We decided not to change the parameter of the displacement gate in layer 4 of the QEC cycle, as its value $\alpha$ contains crucial information about the geometry and the symmetry of the grid code, which could be affected if modifications were allowed.

\begin{table}[hbt]
\begin{tabular}{l*{6}{cccccc}}
\hline
Gate & Parameter  & \quad L1 \quad & \quad L2 \quad & \quad L3 \quad & \quad L4 \quad \\
\hline
$\hat{R}_q$ & $\phi$ & $\pi / 2$ & 0 & 0 & $\pi / 2$ \\
$\hat{R}_q$ & $\theta$ & $\pi / 2$ & $-\pi / 2$ & $\pi / 2$ & $- \pi / 2$ \\
$\hat{ECD}_{qc}$ & $\beta$ & $0.2i$ & $\sqrt{2 \pi}$ & $0.2 i$ & - \\ \hline
\end{tabular}
\caption{Parameters' values optimized for a single half-cycle derived from \cite{Sivak2023}: qubit rotation ($\hat{R}_q (\phi, \theta)$), echoed conditional displacement ($\hat{ECD}_{qc} (\beta)$), cavity virtual rotation ($\hat{VR}_c (\theta_{VR})$).}  \label{table_sBs}
\end{table}

We note here that in our task the required memory times seem to be relatively short, where long-term memory of rare past inputs is not required. For this reason we decided to adopt the recurrent neural network GRU structure, which is easier to train and can also be deployed in an actual experiment since it is able to provide fast inference. Therefore, even though transformers have been very useful in recent applications to physics \cite{vaswani2023, tashev2024}, we discarded them as they do not possess those characteristics that are a plus for the RNNs in the GRU implementation, and make them less suitable for this application.\\
\indent Furthermore, we remind the reader that we train the RNN with a model-based machine learning approach, which is inherently limited to training on simulations, since it requires a physical model of the dynamics. However, we would like to suggest that our approach can be used as pre-training for a neural network that is subsequently tuned with reinforcement learning on experimental data, similar to what has been done experimentally in \cite{Sivak2023} and \cite{Reuer2023}.

\subsection{RNN-agent training} \label{RNNtraining}

For the agent to be included in the QEC scheme as in Fig.~\ref{scheme_QEC} of the main text, we adopted a recurrent neural-network (RNN) with 10 \textit{Gated Recurrent Units} (GRU) cells \cite{cho-etal-2014-learning}, and 2 hidden layers (256, 256), with an activation function $f = tanh()$ in each layer (Table \ref{table_RNN}).
The training was performed over 1000 epochs, with different batch-sizes (6 and 8 for the sBs optimization) for the Monte-Carlo sampling of the measurements giving rise to different trajectories, and collecting the gradient from each of them and applying it to the same network. The learning rate was set to $lr = 10^{-4}$, the biases were all randomly initialized within the range $[-0.1, 0.1]$.\\
For every optimization, multiple agents were trained in parallel in different parallel environments: the best performing agent was then evaluated and post-selected according to the best performance, namely the longer lifetime achieved.\\
For the reward function to maximize we used the fidelity of the final state density matrix $\hat{\rho} = \hat{\rho}(T)$ with respect to the target $\hat{\sigma}$, which is the initial state ($\hat{\sigma} = \hat{\rho} (0)$):
\begin{align}
    \mathcal{F} [\hat{\rho} , \hat{\sigma}] = \left[ \mathrm{Tr} \left( \sqrt{\sqrt{\hat{\sigma}} \hat{\rho} \sqrt{\hat{\sigma}}} \right) \right]^2 \ , \label{fidelity}
\end{align}
which for pure target (initial) states is equivalent to
\begin{align}
    \mathcal{F} [\hat{\rho} , \hat{\sigma}] = \left[ \mathrm{Tr} \left( \sqrt{\hat{\sigma}\hat{\rho}} \right) \right]^2 \ . \label{fidelity_pure}
\end{align}
The choice of this function as a reward to maximize prevents possible artifacts and biases from the initial logical state, as it contains no direct information on the state.

\begin{figure}[h!]
\centering
\includegraphics[width=9cm]{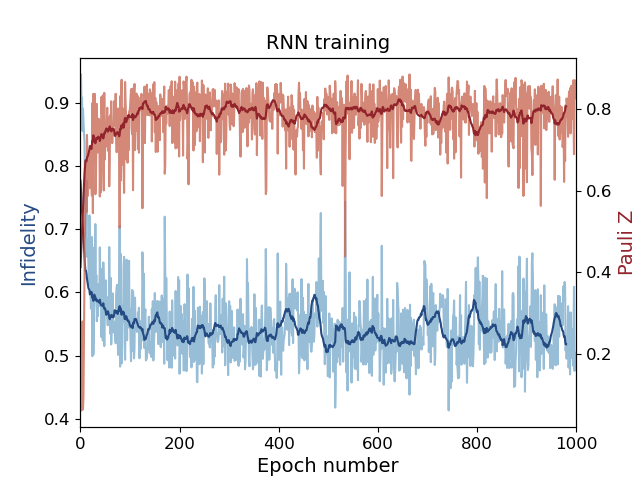}
\caption{Training of the RNN for the NMF QEC strategy. Plot of the final infidelity ($1 - \mathcal{F}$, $\mathcal{F}$ being the fidelity between the final and the initial state) and the Pauli Z expectation value $\langle \hat{Z} \rangle$ as a function of the epoch number during the training with Feedabck-GRAPE. The final fidelity $\mathcal{F}$ has been used as a reward for the agent (see text).}  \label{training}
\end{figure}

An example of the training of a RNN for the optimization of 15 parameters of the sBs QEC protocol is shown in Fig.~\ref{training} for the initial state $| + Z_L \rangle$: the training has been performed over 1000 epochs for trajectories of 10 full sBs QEC cycles, including 2 measurements each.
In blue we show the behavior of the infidelity (defined as $1 - \mathcal{F}$, $\mathcal{F}$ being the fidelity) with respect to the initial state as a function of the training epoch. The dark straight line represents the running average over 40 data points, while the light blue line corresponds to the raw data. The red lines, instead, represent the running average and the raw data of the logical Pauli $\hat{Z}_L$ expectation value, respectively. Since the initial state is the logical $|+ Z_L \rangle$, the maximization of the fidelity via QEC after 10 cycles (or, equivalently, the minimization of the infidelity) corresponds to the maximization of the expectation value of the logical Pauli $\hat{Z}_L$, which is the actual goal of QEC to preserve the logical information contained in it. However, the opposite is not true: maximizing the logical corresponding Pauli $\hat{Z}_L$, in fact, does not lead to maximal fidelity to the initial state, and is not generalizable to other logical states, which is possible using the fidelity, as shown later.
We can also notice that the minimum fidelity is already reached after around 300 epochs, so that in principle the results could have been comparable with a shorter training. However, we decided to opt for a longer training in order for the agent to learn a proper strategy which also minimizes the probability of a drop of logical information, as shown in the Pauli $Z$ expectation value around epoch 1000, in a sort of over-fitting to the problem. \\
\indent Even though we performed the training with only one initial logical state, namely $| + Z_L \rangle$, the use of the state fidelity as a reward function, which is independent of the initial state, guarantees the absence of biases with respect to the choice of the initial state, as explained above. We additionally checked for this effect by comparing the values of the lifetimes for different logical qubits: They turned out to be the same within stochastic sampling fluctuations.

\subsection{RNN-agent performance evaluation and lookup table} \label{Lookup_comparison}
\begin{figure}[h!]
\centering
\includegraphics[width=1.0\linewidth]{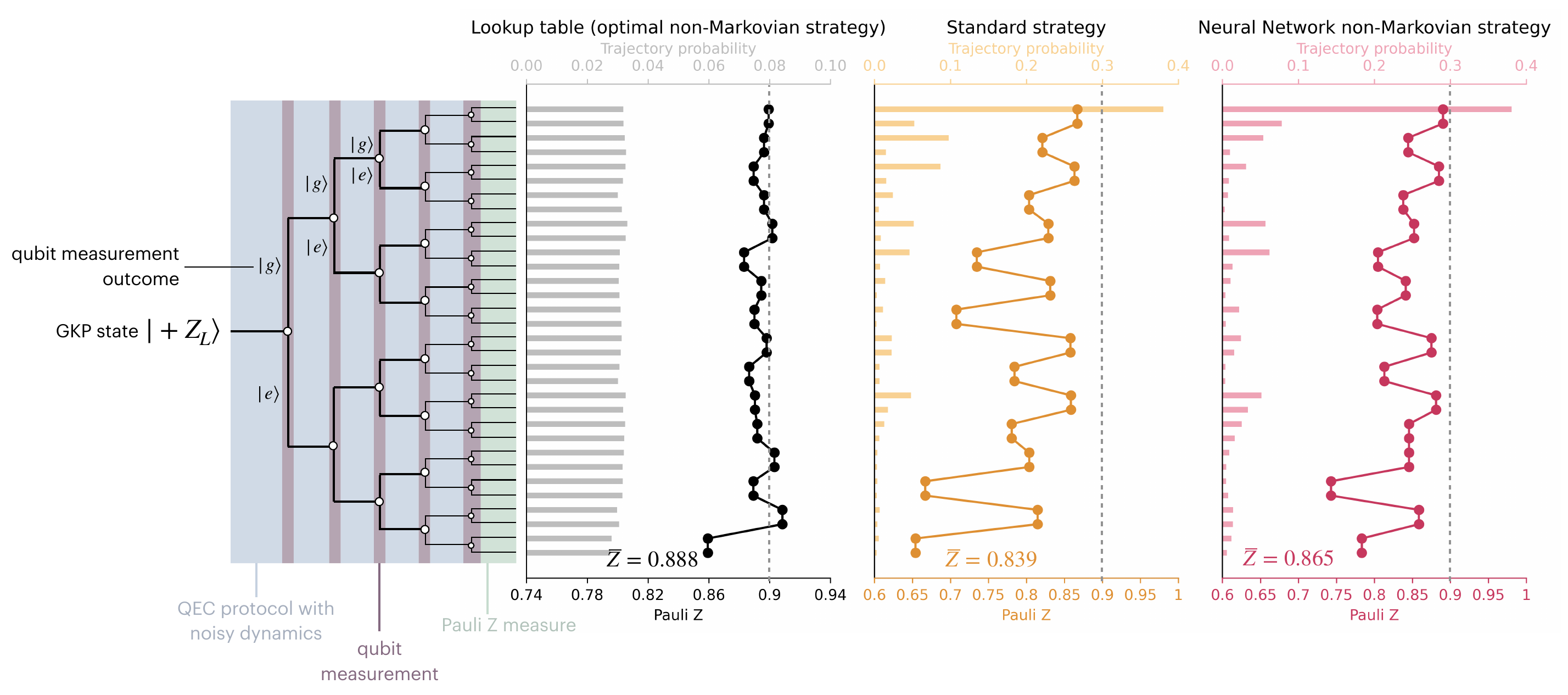}
\caption{Protocol performance comparison with $\Delta = 0.2$. On the left: tree diagram of the quantum error correction trajectories generated by the different transmon measurement outcomes. Each grey area represents the correction protocol with noisy dynamics, the purple areas correspond to the ancillary qubit measurements, and the final green area is the final Pauli Z operator evaluation (perfect). On the right: histogram of each trajectory probability (scale on the top axis) and plot of the Pauli Z expectation value (bottom x axis) for the corresponding final state, for the optimal strategy (left), the standard strategy (center) and our non-Markovian feedback strategy based on the recurrent neural network (RNN). Note that the scale of the probability distribution provided in the top x axis is different among the plots, as well as the one of the bottom x axis for the final Pauli Z value.} \label{performance}
\end{figure}

\begin{figure}[h!]
\centering
\includegraphics[width=1.0\linewidth]{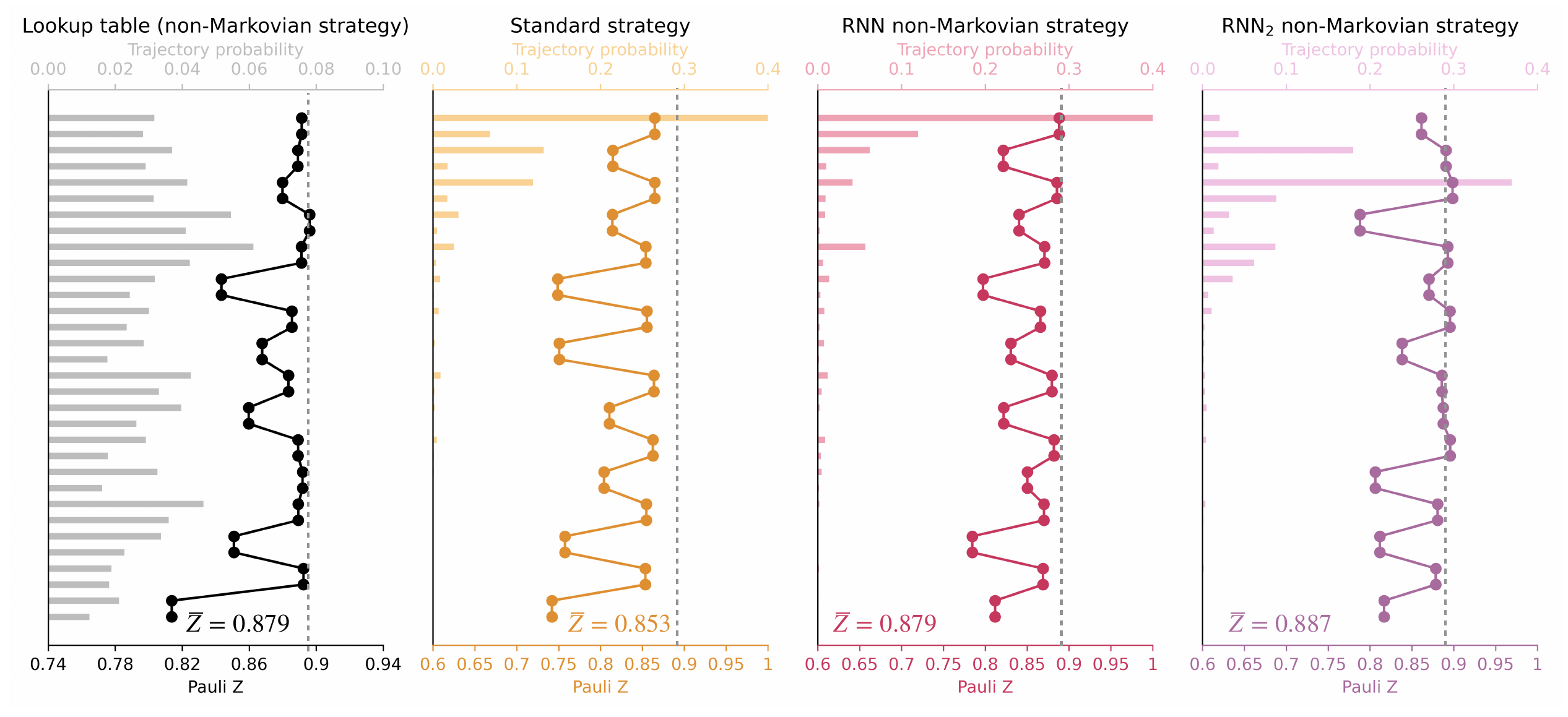}
\caption{Protocol performance comparison with $\Delta = 0.34$ without retraining from the values for $\Delta = 0 .2$. Histogram of each trajectory probability (scale on the top axis) and plot of the Pauli Z expectation value (bottom x axis) for the corresponding final state, for the optimal strategy (left), the standard strategy (center) and two non-Markovian feedback strategies based on recurrent neural networks (RNNs) with different training parameters. Note that the scale of the probability distribution provided in the top x axis is different among the plots, as well as the one of the bottom x axis for the final Pauli Z value.} \label{performance2}
\end{figure}

In order to evaluate the performance of our neural network and compare it against the idealized best case, we built a lookup table of parameters (with a different set of parameters for each sequence of measurements outcomes). This guarantees in principle that the perfect strategy (within the bounds set by the ansatz for the gate sequence) could be found, albeit under the constraint of a finite duration of the sequence: the lookup table grows exponentially with time. We then optimized these parameters using feedback-GRAPE algorithm described in section \ref{theoryFG}, with the goal to optimize the final state fidelity averaged over all the possible measurement results. This represents by construction the upper limit threshold for the parameters of the sBs protocol of a given model. Then, we compared the logical qubit lifetime obtained with this lookup table, with the standard strategy and with our neural network-based non-Markovian strategy. We highlight here that the neural network used 
The results for the $| + Z_L \rangle$ initial state and $\Delta = 0.2$ are presented in Fig.~\ref{performance}: they show the probability distribution (as an histogram) and the corresponding logical value (as a dot-line plot) after two full QEC cycles for each trajectory formed by any possible qubit measurements outcomes, comparing the optimal strategy (Fig. \ref{theoryFG}, left, in black), the standard strategy (Fig. \ref{theoryFG}, center, in yellow) and the non-Markovian strategy based on a RNN as presented in this work (Fig. \ref{theoryFG}, right, in red). We adopted here the 'low' noise model as described in the main text and the section \ref{generalNoise}, according to experimental data from \cite{Sivak2023}. 
While the lookup table optimal parameters reaches a logical value which is 94.0$\%$ of the optimal one (obtained with the lookup table), while our non-Markovian strategy improves it to 97.0$\%$. This means a reduction by a factor 2 of the logical error for two full QEC cycles, from 6.0 $\%$ to 3.0$\%$. \\
However, there are two severe limitations inherent in the lookup table approach: the first is the generalizability to longer times, and the second is the generalizability to different parameters. The former is extremely challenging due to the exponential increase of the number of parameters of the lookup table, as well as the computational time to calculate and the memory to store them. This prevents the lookup table approach to be used in realistic QEC situation with many cycles. The latter, instead, makes this approach limited to a very narrow interval of experimental parameters, making it more difficult to be implemented in a real experiment when the corresponding theoretical model is not very accurate.\\
\indent In order to exclude possible limitations of the neural network architecture itself and to find evidence of the generalizability limitations of lookup tables, we consider the same set of parameters obtained for a GKP model with $\Delta = 0.2$ and, without retraining or further optimization, we utilize them in a quantum error correction protocol for a state with a different squeezing, namely $\Delta = 0.34$. In Fig.~\ref{performance2} we show the comparison for this setup of the performance of the lookup table, the standard strategy, and the neural network-based non-Markovian feedback approach with two different training of the neural network (for more details see Section \ref{RNNagents}). 
In this situation, while the lookup table parameters provide better performance than the standard approach, highlighting the necessity of an adaptive strategy, the two neural networks have a performance comparable and higher, respectively, than the lookup table, without retraining. This is an evidence of the higher generalizability of the neural networks with respect to the static lookup table. 
\\
All in all, this shows that our neural network approach is nearly optimal with respect to the lookup table approach, with the additional significant advantages of scalability to arbitrary long times and robustness against perturbations of the theoretical model. In addition, while the probability of each trajectory when using the optimal parameters of the lookup table is almost uniform, the probability distribution is not significantly changed from the standard to the neural network non-Markovian approach. This, as highlighted in \cite{Sivak2023}, turns out to be beneficial from the experimental point of view in order to preserve the accuracy of the qubit measurements. 

\section{Additional numerical results} \label{Results}

\subsection{Autonomous quantum error correction} 

\begin{figure}[h!]
\centering
\includegraphics[width=1.0\linewidth]{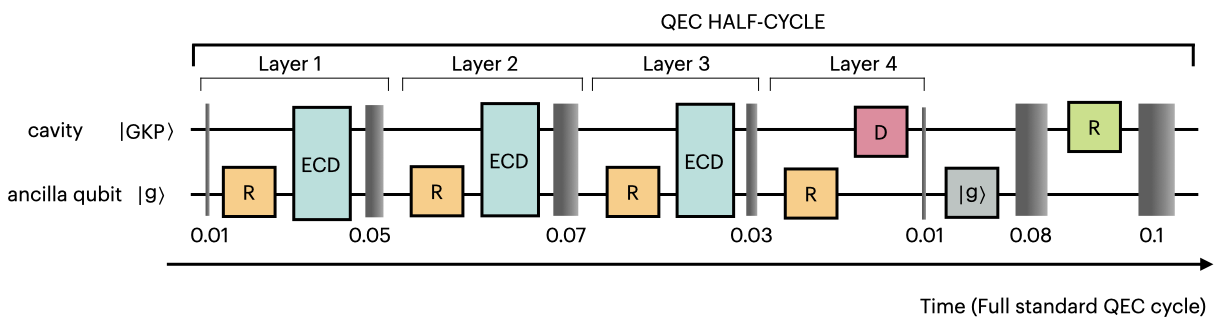}
\caption{Autonomous QEC scheme. Scheme of the half-cycle sBs autonomous QEC, without measurements. The ancillary qubit is still reset after layer 4. The grey vertical strips represent the dynamics duration referred to the length of a full standard QEC as in Fig.~\ref{scheme_QEC}: the summation returns 0.35 instead of 0.5 because the dynamics for the qubit reset has been reduced from 0.23 to 0.08 due to the absence of the measurement and readout. This means that a full cycle of the autonomous QEC lasts 0.7 of the measurement-feedback one including measurements.} \label{scheme_auto}
\end{figure}

Recently it has been shown in an experiment that it is possible to perform autonomous quantum error correction on the GKP code with the sBs protocol, resulting in a longer qubit's lifetime with respect to the vanilla measurement-based sBs protocol \cite{AutonomousQEC2024}. 
We considered the autonomous QEC scheme as represented in Fig.~\ref{scheme_auto}, realistically reducing the time dynamics for the reset from $0.23$ to $0.08$ expressed in units of the full measurement-based QEC cycle, $\tau_{cycle}$. As a consequence, the duration of half-cycle for the autonomous QEC will be $0.35$ instead of $0.50$ as for the measurement-based QEC half-cycle. This value has been used as a realistic estimate resulting from the absence of the measurement process. As a consequence, the ancillary qubit reset happens in the autonomous QEC $\approx 43 \%$ more times than for the measurement-based sBs protocol. 

\subsection{Average channel fidelity: evaluation and lifetime}
\begin{figure}[h!]
\centering
\includegraphics[width=12cm]{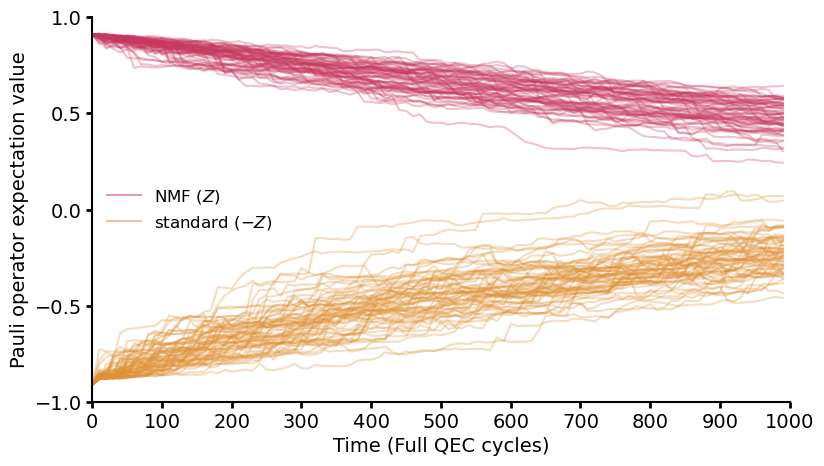}
\caption{Trajectories' sampling. Trajectories (each of them averaged over a batch of sampled trajectories) of the logical Pauli operator expectation value as a function of time, when applying the standard and the NMF protocol, respectively. For the standard strategy the value $- \langle \hat{Z} \rangle$ is shown for better readability.} \label{sampling} \end{figure}

The average trajectories of the Pauli operators' expectation values, which are shown e.g. in Fig.~\ref{results}(a) in the main text, are calculated from $6 \times 85$ (for NMF approach) and $8 \times 64$ (for the standard protocol) raw trajectories, where the first number is the batch-size of the sampling and the second number is the repetition number for each of them. In Fig.~\ref{sampling} we show these raw trajectories (from which Fig.\ref{results}(a) is derived) which already contain the average over the batch-size. The standard deviation shown in the plots is calculated on these trajectories as well.
From those averages, we were able to extract the lifetimes $T_k$ for each initial logical state $| k_L \rangle$ by fitting the trajectories with the exponential function $\langle \hat{P}_k \rangle (t) = \langle \hat{P}_k \rangle (0) \ e^{-t/T_k}$, with Pauli operators $P_{k = X, Y, Z}.$

\begin{figure}[h!]
\centering
\includegraphics[width=12cm]{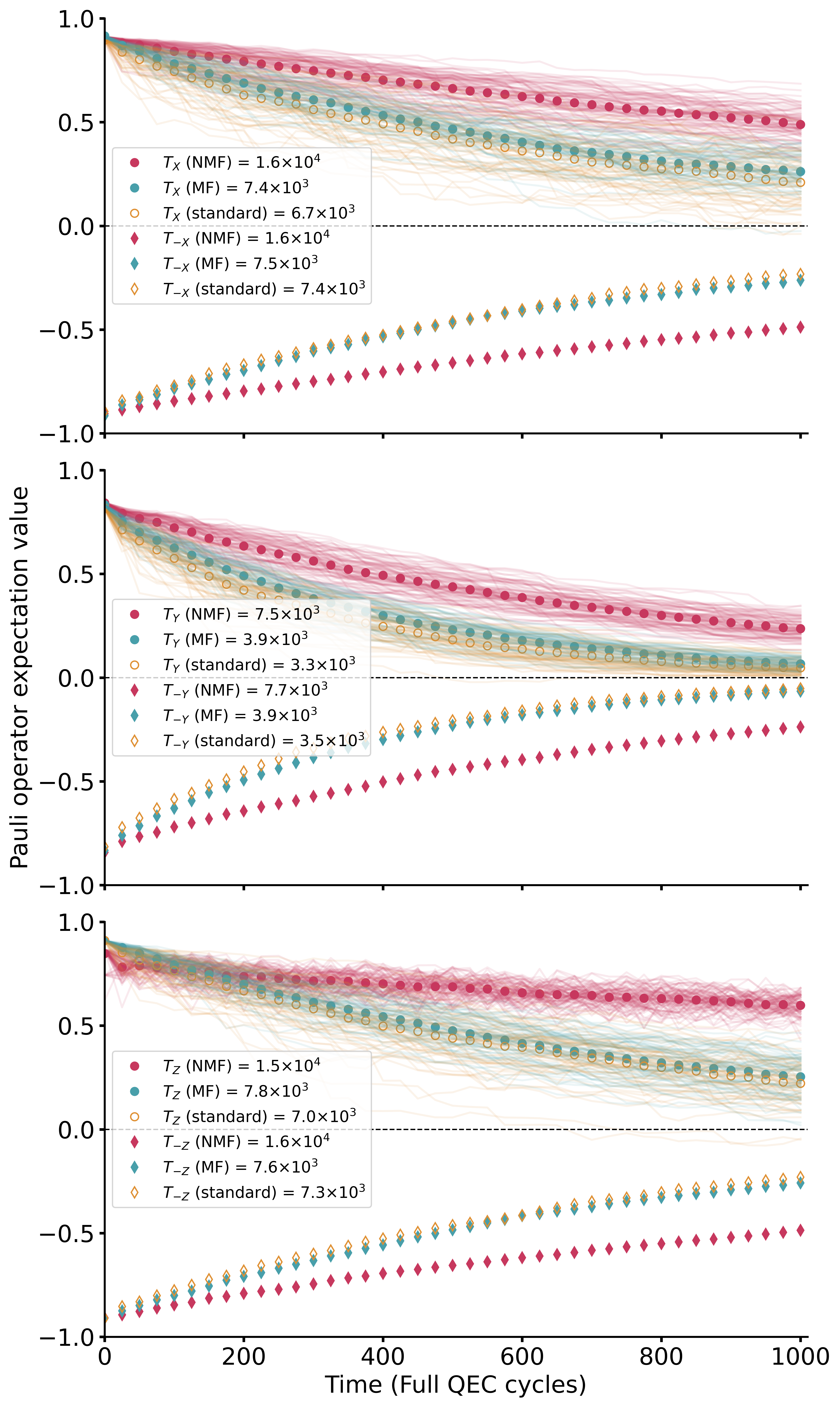}
\caption{Time evolution of logical Pauli eigenstates. We show the Pauli operator expectation values for the two eigenstates of each of the Pauli logical operators $\hat{X}$ (top), $\hat{Y}$ (center) and $\hat{Z}$ (bottom), respectively. For each of them we compare the results for different QEC protocols, namely the non-Markovian feedback (NMF, red), the Markovian feedback (MF, blue) and the standard strategy (standard, yellow).} \label{Pauli_all} 
\end{figure}

\begin{figure}[h!]
\centering
\includegraphics[width=12cm]{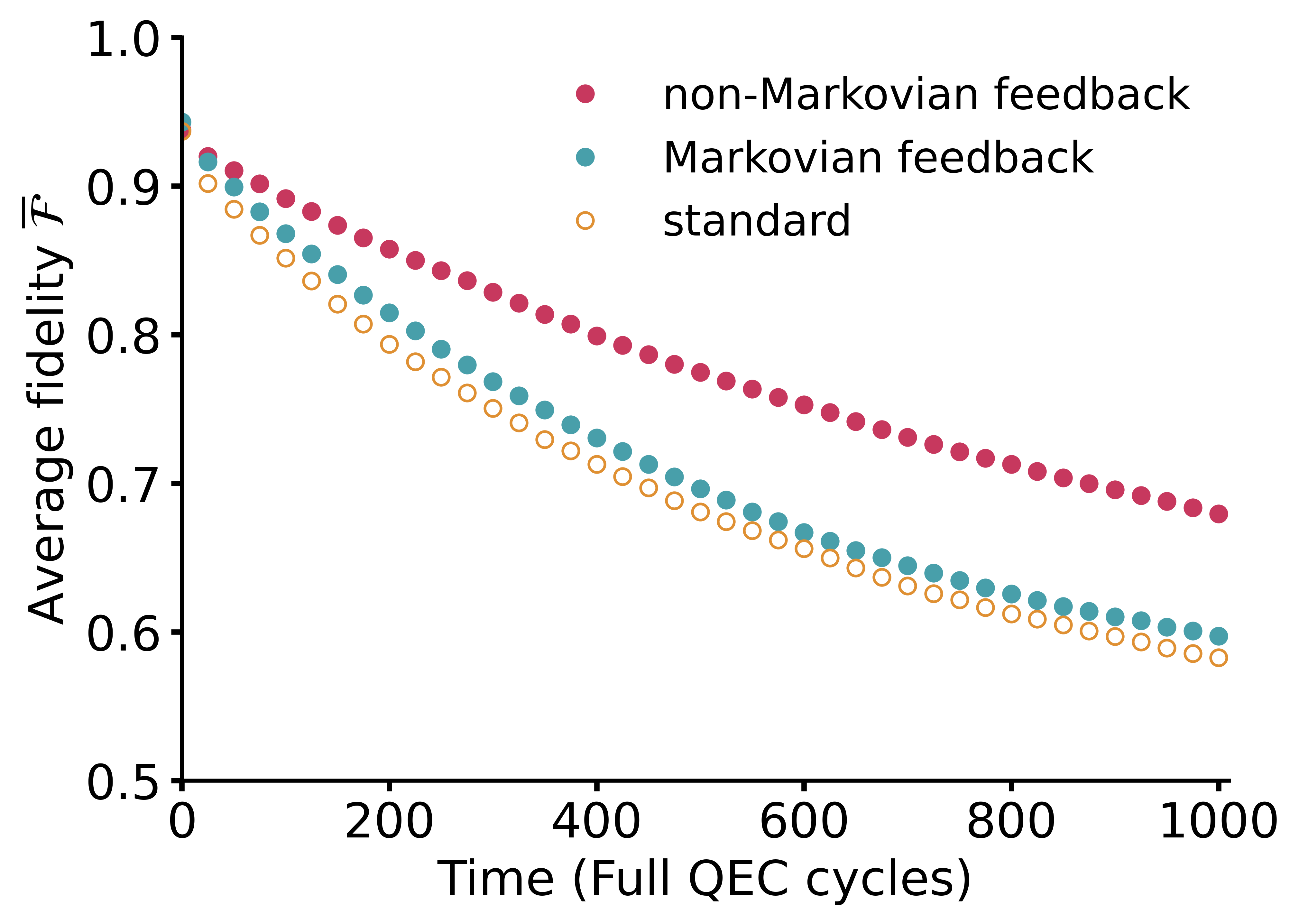}
\caption{Time evolution of the average channel fidelity. We show the time evolution of the average channel fidelity over six Pauli eigenstates for different QEC protocols, namely the non-Markovian feedback (NMF, red), the Markovian feedback (MF, blue) and the standard strategy (standard, yellow).} \label{Channel_fidelity_all} 
\end{figure}

Following \cite{Sivak2023}, we can introduce the average channel fidelity, which is defined as 
\begin{align}
\overline{\mathcal{F}} (t) = \dfrac{1}{2} + \dfrac{1}{12} \bigg\lbrace \mathrm{Tr} \lbrace \hat{X} \hat{\rho}_X (t) \rbrace + \mathrm{Tr} \lbrace \hat{Y} \hat{\rho}_Y (t) \rbrace + \mathrm{Tr} \lbrace \hat{Z} \hat{\rho}_Z (t) \rbrace - \mathrm{Tr} \lbrace \hat{X} \hat{\rho}_{-X} (t) \rbrace - \mathrm{Tr} \lbrace \hat{Y} \hat{\rho}_{-Y} (t) \rbrace - \mathrm{Tr} \lbrace \hat{Z} \hat{\rho}_{-Z} (t) \rbrace \bigg\rbrace \,,
\end{align}
as shown in Fig.\ref{Channel_fidelity_all} for the standard strategy, as well as for the Markovian and non-Markovian feedback approaches.
Due to the symmetric behavior of the Pauli eigenstates with opposite eigenvalues, for our system the average channel fidelity is simply given by \cite{Sivak2023}
\begin{align}
\overline{\mathcal{F}} (t) = \dfrac{1}{2} + \dfrac{1}{6} e^{-t/T_X} + \dfrac{1}{6} e^{-t/T_Y} + \dfrac{1}{6} e^{-t/T_Z} \,,
\end{align}
where $T_{k = X, Y, Z}$ are the lifetimes extracted from the decays of the logical states. The average channel fidelity shown in Fig.~\ref{results}(b) in the main text, as well as the ones in Fig.~\ref{lifetimes_all}, is calculated as $T = \dfrac{3}{\frac{1}{T_{-X}} + \frac{1}{T_{-Y}} + \frac{1}{T_Z}}$, since $T_{-X} = T_X$ and $T_{-Y} = T_Y$ hold. In this way we also verify that $T_X = T_Z$ within the variation due to the sampling of a limited number of trajectories (500+).\\
We provide for completeness in Fig.~\ref{Pauli_all} the average trajectories for all the six Pauli operators' eigenstates for the standard, the Markovian and non-Markovian feedback strategies, respectively. It is possible to notice the symmetry between the positive and the negative valued eigenstates within numerical fluctuations, and in particular the outperformance of the non-Markovian feedback approach with respect to the others. 

In order to provide a comparison of the efficiency of different QEC strategies, we utilize the entanglement fidelity $\mathcal{F}_e$, which is defined in \cite{Sivak2023} for the GKP code as
\begin{align}
\mathcal{F}_e = \dfrac{3 \overline{\mathcal{F}} - 1}{2} \,.
\end{align}
For the full cycle duration of $\tau_{\text{cycle}} = 10 \ \mu$s, cavity decay rate $\kappa = 1 / (610 \ \mu \text{s})$ and the logical lifetimes extracted from the average trajectories, we obtain the entanglement infidelities
\begin{align}
\left( 1 - \mathcal{F}_e \right)_{\text{std}} = 1.5 \times 10^{-3} \,, \\
\left( 1 - \mathcal{F}_e \right)_{\text{NMF}} = 4.3 \times 10^{-4} \,,
\end{align}
for the standard and the non-Markovian feedback strategies, respectively. These values can be compared with the theoretical threshold provided in \cite{Performancebosonic2018} for a value of the dimensionless decay rate $\gamma = 1 - e^{- k \tau_{\text{cycle}}} = 0.016$, which is $\left(1 - \mathcal{F}_e \right)_{th} \approx 10^{-6}$ for an average photon number $\langle \hat{n} \rangle \leq 5$ as in our case. However, this theoretical threshold is computed in the ideal case, assuming an optimal recovery operation and not considering the noise occurring during the correction protocol, nor the ancillary transmon noise used to control the cavity \cite{Sivak2023}. We highlight that this latter represents the main limitation of the quantum error correction protocol efficiency, as mentioned in \cite{Sivak2023}, and it is included in our simulations to faithfully capture the physics of real experiments. Therefore, while this comparison can be helpful to benchmark the technical limits of the state of the art error correction, unfortunately it cannot be a faithful value to evaluate the optimality of our strategy.

\begin{figure}[h!]
\centering
\includegraphics[width=12cm]{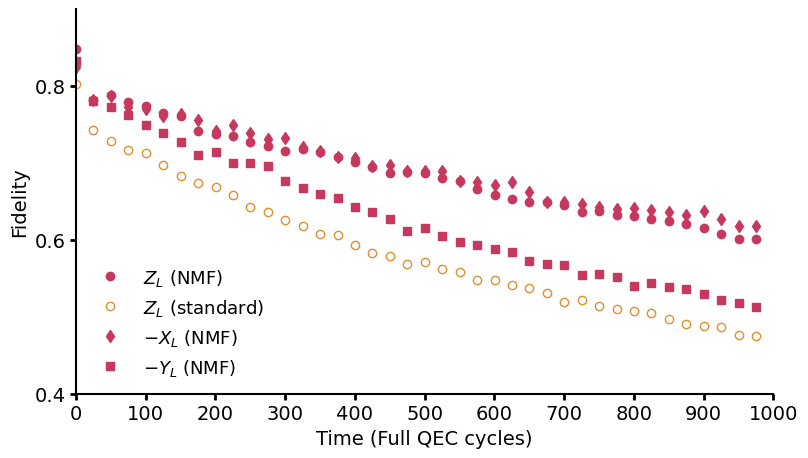}
\caption{Time evolution of the state fidelity of the GKP code with respect to the initial state when performing QEC according to the sBs protocol without (standard) and with non-Markovian feedback. The values displayed refer to different initial logical states, namely $Z_L$, $-X_L$ and $- Y_L$.} \label{fidelity_traj}
\end{figure}

It is also interesting and insightful to have a look at the behavior of the state fidelity during the QEC process. In fact, our neural-network agents are given as a reward, i.e. function to maximize, the final fidelity $\mathcal{F}$ of the density matrix of the state with respect to the initial state $| +Z_L \rangle$, as defined in Eq.~(\ref{fidelity_pure}). However, in order to evaluate the QEC performance, the different agents and their strategies are evaluated on the logical qubits' lifetime, namely calculating the time-evolution and the decay rate of the logical Pauli operators. 
In Fig.~\ref{fidelity_traj} we show the fidelity as a function of the QEC cycle number for different logical states for the standard and the best NMF strategies. Similarly to the case of the logical lifetime, even the state fidelity remains higher over time with the time-dependent approach with memory compared with the standard sBs protocol.

\subsection{RNN agents and post-selection} \label{RNNagents}
\begin{figure}[h!]
\centering
\includegraphics[width=0.65\linewidth]{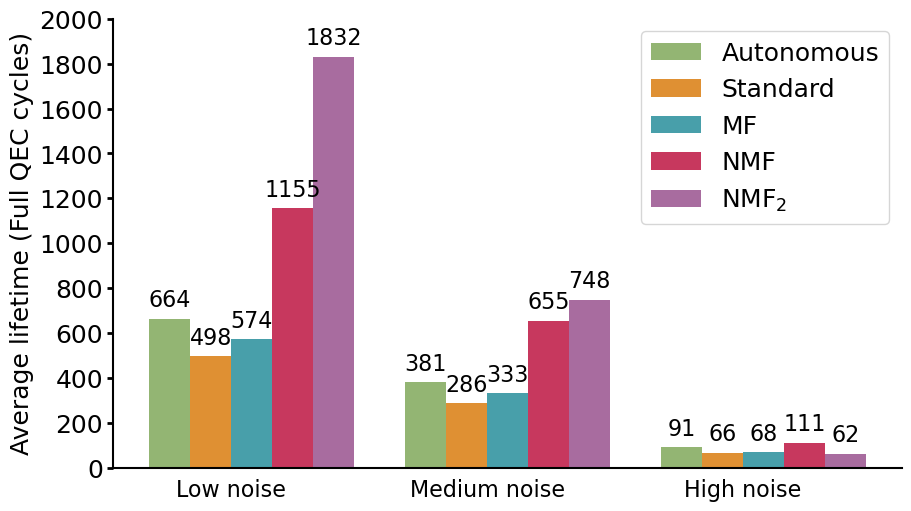}
\caption{Average channel fidelity of different strategies. Comparison of the average lifetime for different strategies and noise levels as defined in Table \ref{table_noise}.} \label{lifetimes_all}
\end{figure}

The results shown for the NMF approach have been obtained by training multiple (20) RNN agents, evaluating all of them and post-selecting the one which was offering the highest performance in terms of logical lifetime's extension. The results shown in the main text were obtained with the best agent among the ones trained using the dynamics shown in Fig.~\ref{scheme_QEC} and high noise level, with an initial physical GKP state $| Z_L \rangle$ with $\Delta = 0.2$. \\
We will discuss about the generalizability of these agents in the following sections. However, as it was evident even from the main results, the agent trained with given dynamics and noise still outperforms the standard strategy just as well even for different circuit duration and noise level. Despite this, we also evaluated in the same conditions another agent based on the same RNN model, that we called NMF$_2$, which was post-selected after training on a simplified dynamics (as depicted in Fig.~\ref{simple_dynamics}) and high noise level. Its performance is shown in Fig.~\ref{lifetimes_all} in comparison with all other strategies shown also in Fig.~\ref{results}(b) in the main text. As it is visible, it outperforms by a factor $\approx 1.7$ even the other RNN with NMF in the low noise level. However, this gain decreases for medium noise and at high noise it even fails to perform even just as well the standard sBs protocol. The reason of this reduced generalizability is found in the different probability distribution of the ancillary measurements' outcomes (Fig.~\ref{strategy_2}), as discussed in the next sections. For this reason, and because of its lower adaptability to a real experimental setup, we decided to discard this strategy from the main results.

\subsection{Generalization to different noise levels} \label{generalNoise}

In our work we considered three different noise levels, reflecting real values of recent experiments: namely, in our numerical results we simulated the dissipative dynamics of the cavity, given by the cavity relaxation time ($T_s$), and the qubit relaxation ($T_1$) and dephasing ($T_2$). For the high noise level we used the values in Ref.~\cite{CampagneIbarcq2020}, while for the low level we referred to the more recent Ref.~\cite{Sivak2023}. The medium noise level was obtained by simply doubling the times of the high level. These values are represented in Table \ref{table_noise}, with the times expressed in units of duration of a full sBs QEC cycle, which similarly to Ref.~\cite{Sivak2023} we set to $\tau_{\text{cycle}} = 10 \ \mu$s.

\begin{table}[h]
\begin{tabular}{l*{3}{ccc}}
\hline
Noise level & \quad $T_s / \tau_{\text{cycle}}$ & \quad  $T_1 / \tau_{\text{cycle}}$ & \quad $T_2 / \tau_{\text{cycle}}$ \\
\hline
Low \cite{Sivak2023} & \quad 61 \quad & \quad 28 \quad & \quad 23.8 \\
Medium & \quad 49 \quad & \quad 10 \quad & \quad 12 \quad \\
High \cite{CampagneIbarcq2020} & \quad 24.5 \quad & \quad 5 \quad & \quad 6 \quad \\ \hline
\end{tabular}
\caption{Noise levels. Table of the components' lifetimes used for the dynamics, in units of a full standard sBs cycle's duration, provided in Fig.~\ref{scheme_QEC}, which is $\tau_{\text{cycle}} = 10 \ \mu$s. $T_s$ is the cavity relaxation time, $T_1$ and $T_2$ are the ancillary qubit's relaxation and dephasing times, respectively.}  \label{table_noise}
\end{table}

In particular, as we can notice in Fig.~\ref{results}(b) of the main text and in Fig.~\ref{lifetimes_all}, the logical qubit's lifetime decreases as the noise level increases, but the ratio is not linear: in fact, by doubling the components' lifetime from high to medium, the GKP code's lifetime increases by a factor higher than 2. In particular, this value depends on the strategy adopted: by decreasing the noise, the standard strategy increases by a factor $\approx 4$, while the NMF approach provides an improvement of almost a factor $\approx 6$.

\subsection{Generalization to different dynamics}
\begin{figure}[h!]
\centering
\includegraphics[width=1.0\linewidth]{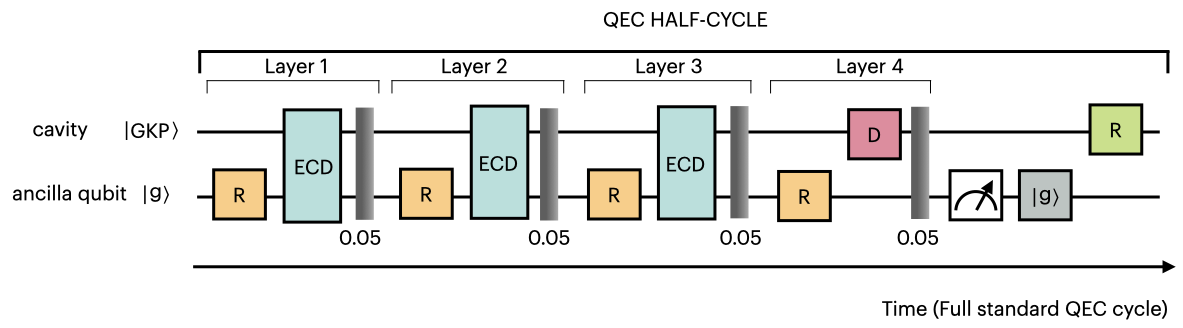}
\caption{Simplified dynamics QEC scheme. Scheme representing the sBs QEC protocol similarly to Fig.~\ref{scheme_QEC}, with a simpler dynamics scheme, represented by the grey vertical strips. In this case, we adopted a constant duration of the dynamics of each layer, and instantaneous measurement and reset (i.e.: no time evolution for the measurement and the ancillary qubit reset).}  \label{simple_dynamics}
\end{figure}

\begin{figure}[h!]
\centering
\includegraphics[width=0.65\linewidth]{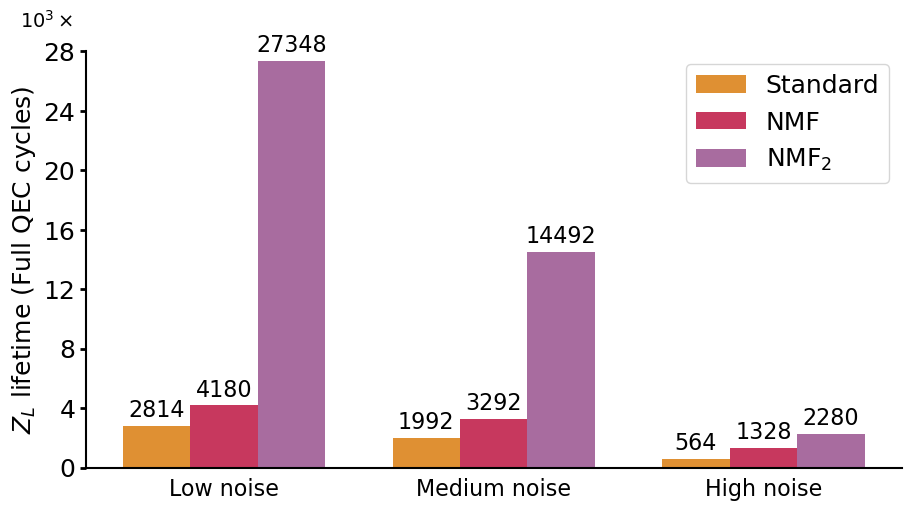}
\caption{Comparison of the lifetimes of the $Z_L$ state for different sBs protocols with the dynamics simulation in Fig.~\ref{simple_dynamics}. The standard sBs protocol uses the constant optimal gates' parameters, while NMF and NMF$_2$ use the non-Markovian-feedback approach, and they were trained on different dynamics (see text).} \label{simple_lifetime}
\end{figure}

One strength of our NMF approach based on recurrent neural network is its versatility and generalizability to different noise and dynamics, which proves the presence of an underlying common feature for the quantum error correction of the GKP code. In fact, the same neural network agent trained for a given dynamics model and noise level, can perform just as well for a different combination of dynamics and noise, even if subtle remarks have to be raised. \\
\indent Since our simulations are not limited to real experimental conditions, we can show how the results would change under idealized conditions. As an example, we considered a simplified dynamics scheme that includes instantaneous measurement and qubit reset, and equal time duration of each layer of the sBs protocol, as shown in Fig.~\ref{simple_dynamics}. In particular, we compared the standard protocol, as well as two RNN with non-Markovian feedback, namely NMF and NMF$_2$ (this latter described in the previous section): the former was trained on the original dynamics of Fig.~\ref{scheme_time}, while the latter on the simple dynamics of Fig.~\ref{simple_dynamics} (and both on high noise level). The lifetime of the $Z_L$ logical state is shown in Fig.~\ref{simple_lifetime} for the different strategies and for three different noise levels (as described in Table \ref{table_noise}). As we can notice, both the NMF strategies outperform the standard protocol, confirming that the memory is a widespread advantage for improving QEC, independent of the circuit dynamics. However, we also do observe that the agent trained with the same circuit dynamics outperforms the standard strategy by almost a factor $\approx 10$. The reason can be found in the specific characteristics of this circuit, where the measurement ratio is higher (because the overall cycle duration is lower) and no dissipation is present during measurement and reset. This leads to a better error correctability, which makes not useful to have a high probability to measure the ancillary qubit in its ground state $g$, unlike in the realistic case. As a consequence, the agent trained on this simplified dynamics, NMF$_2$, optimized the parameters of the gates in disregard of the measurement outcome, as it is visible in the section on the strategy analysis. In fact, as the noise level increases, the advantage of such an approach becomes less evident: with high noise, the NMF$_2$ outperforms the standard strategy by a factor $\approx 4$. This behavior is similar for the evaluation on a realistic dynamics (see Fig.~\ref{lifetimes_all}), even though in that case the NMF$_2$ agent is failing even to perform as good as the standard strategy in the high noise conditions. 

\begin{figure}[t]
\centering
\includegraphics[width=0.5\linewidth]{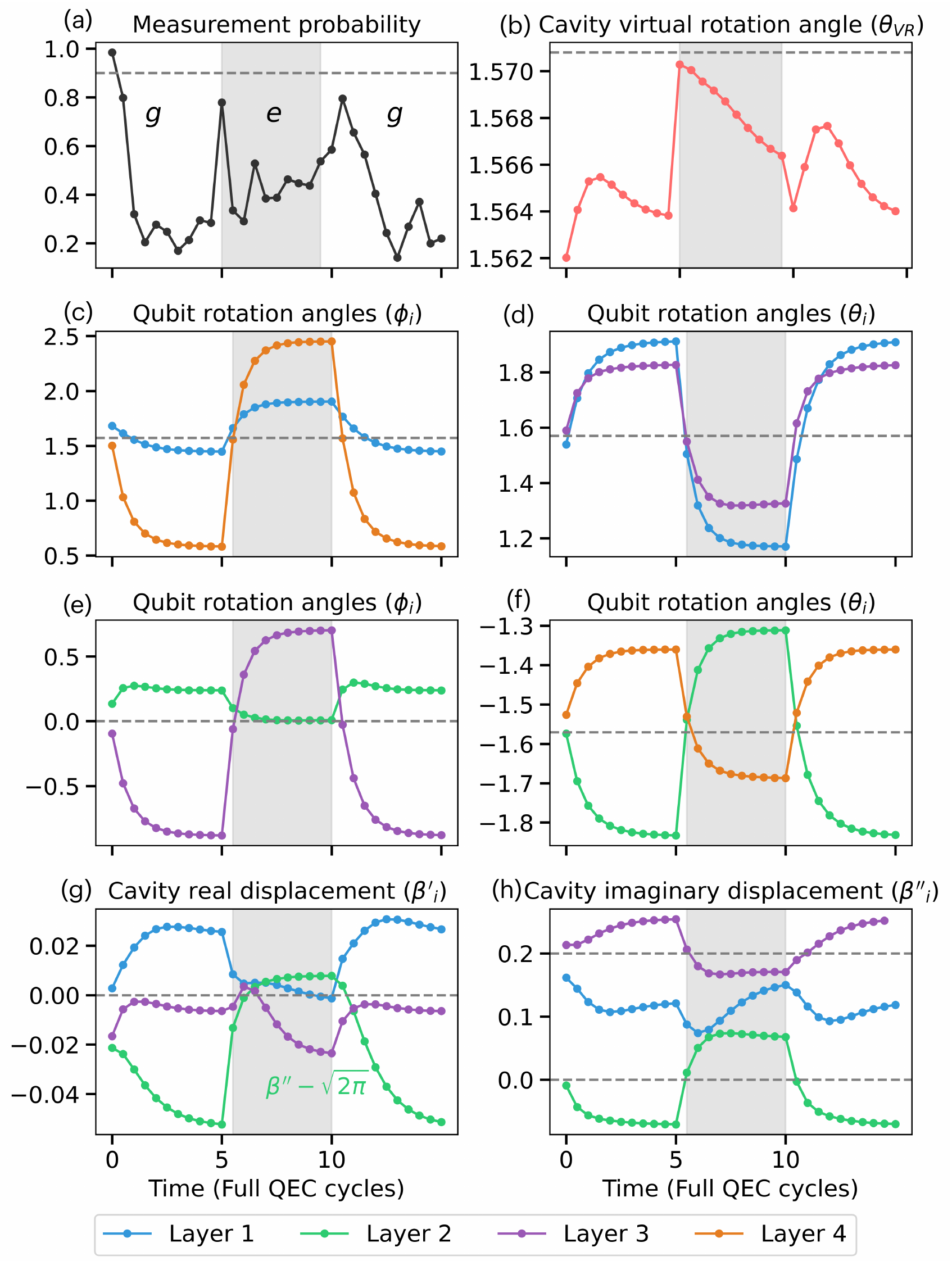}
\caption{Optimized time-dependent parameters for the RNN of the NMF$_2$ approach, evaluated with the dynamics in Fig.~\ref{scheme_time} with low noise rate. The plot corresponds to the one in Fig.~\ref{strategy} for NMF, fixed deterministic measurements' outcomes.} \label{strategy_2}
\end{figure}

\subsection{Strategy analysis}
\begin{figure}[t]
\centering
\includegraphics[width=0.5\linewidth]{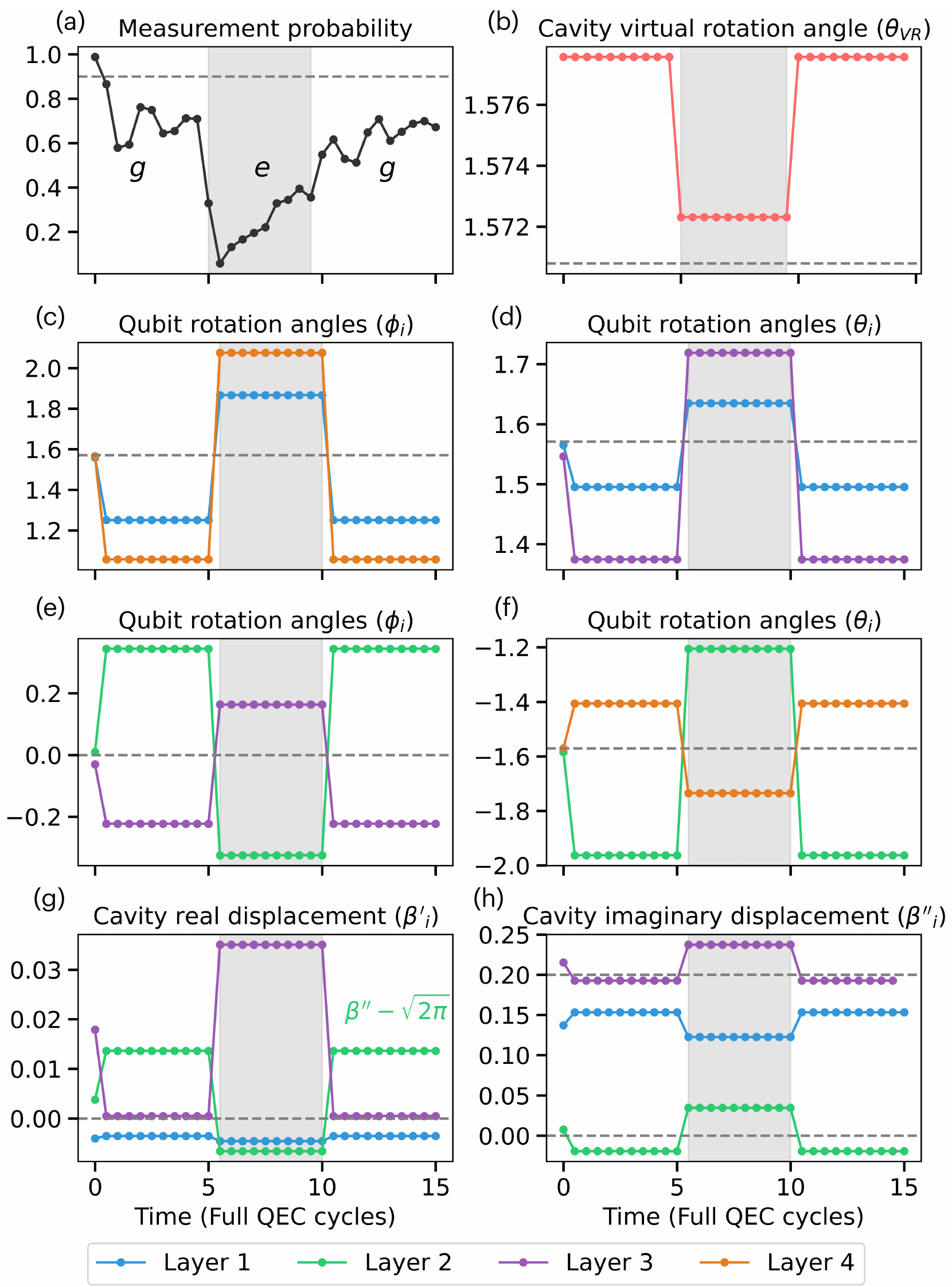}
\caption{Optimized time-dependent parameters for the feedback agent based on feed-forward neural network without memory (MF), evaluated with the dynamics in Fig.~\ref{scheme_time} with low noise rate. The plot corresponds to the one in Fig.~\ref{strategy} for NMF, fixed deterministic measurements' outcome.} \label{strategy_FNN}
\end{figure}

We analyze here more in detail the strategies that the different RNNs find when trained in different conditions. In particular, we observe that all the recurrent neural networks show a characteristic time dependence of the output parameters, which behaves as an exponential saturation to a certain value. This is true for both sequences of $g$ as well as $e$ measurements' outcomes, but with different saturation levels. Due to this behavior, we can fit the parameters' values after a $g/e$ qubit's measurement's outcome as 
\begin{align}
    \pi_{g/e} (t - \overline{t}) = \pi_{e/g (0)}(t = \overline{t}) \ e^{- \gamma_{g/e} (t - \overline{t})} + \pi^\infty_{g/e} \left( 1 - e^{-\gamma_{g/e} (t - \overline{t})} \right)  \ ,
\end{align}
where $\overline{t}$ is the first time-step at which the sequence of identical $g/e$ measurements started, $t$ is the current time-step, $\pi^\infty_{g/e}$ is the saturation level for a theoretical infinite number of consecutive $g/e$ measurements' outcomes, $\gamma_{g/e}$ is the parameter decay rate for consecutive $g/e$ measurements. The initial value of one parameter for the very first step is $\pi_0 (0)$. Then, if we have consecutive $g$ measurements' outcomes, the neural network adjusts the outputs to $\pi_{g} (t) = \pi_0(0) \ e^{- \gamma_{g} t} + \pi^\infty_{g} \left( 1 - e^{-\gamma_{g} t } \right)$. Ideally, if we measure only $g$ on the ancillary qubit, we would end up with the values $\pi_{g} (t \rightarrow \infty) = \pi^\infty_{g}$. If we get a sequence of 5 times $g$ we will get the parameter $\pi_{g} (5) = \pi(0) \ e^{- 5 \gamma_{g} } + \pi^\infty_{g} \left( 1 - e^{-5 \gamma_{g}} \right) \approx  \pi^\infty_{g}$. However, if we get a measure $e$ after a sequence of 5 times $g$, then the next parameter's value will be $\pi_{e} (1) = \pi_g(5) \ e^{- \gamma_{e} } + \pi^\infty_{e} \left( 1 - e^{-\gamma_{e} } \right)$, where $\pi_g(5)$ was the last value used for the $g$ measurement, as calculated before. All in all, however, we realize that the most important term of this expression which determines whether a strategy is failing in QEC or not, is the saturation value $\pi^\infty_{g/e}$, even if the time dependence with an exponential saturation is necessary to outperform. Moreover, it is important to notice that the optimal parameters' choice for a given experimental setup (circuit duration and noise level) is not unique, as there is some degree of freedom due to symmetry in the choice of the rotation angles and conditional displacement's parameters. \\
Another relevant aspect to be analyzed is the probability measurement's outcome determined by the noise level and, in particular, by the parameters' values chosen. We can then compare the strategy of the NMF shown in Fig.~\ref{strategy} in the main text, and the one NMF$_2$ in Fig.~\ref{strategy_2}: despite having different signs in the exponential saturation behavior, which is part of the degree of freedom mentioned above, the main difference is the amplitude of those values. This is reflected into the measurement's probability as shown in the top left panel of Fig.~\ref{strategy_2} and Fig.~\ref{strategy} in the main text: while the NMF strategy returns a probability for the $g$ measurement $p(g) = 0.9$ constant as long as $g$ are indeed measured, for the NMF$_2$ this probability drops to $p(g) \approx 0.3$, making more frequent the measurement of the qubit in the $| e \rangle$ state. Although this has no negative consequence here, except the limitation of generalizability to very different noise levels, this can be detrimental in experiments, where it has been shown that increasing the probability to measure $e$ can lead to more errors \cite{Sivak2023}.\\
When using a feed-forward neural network, instead, the output parameters can only assume three possible values, as there is no memory and therefore only a Markovian response: namely, $\pi_0$ as initial value, $\pi_g$ in case of $g$ measurement outcome, and $\pi_e$ in case of $e$ measurement (Fig.~\ref{strategy_FNN}). Moreover, as we can notice, the values $\pi_g$ and $\pi_e$ assume symmetrically opposite values with respect to the initial value $\pi_0$.

\subsection{Injected displacement errors}
In Section \ref{noise} we have provided a description of the quantum noise that the system of the oscillator and the coupled ancillary transmon is subject to. The noisy dynamics, included in our simulations, affects both the cavity and the qubit during the implementation of the gates to realize the quantum error correction protocol. As a result, the system acts as a quantum memory with an initial logical state $| L \rangle$, and the QEC protocol corrects the cavity loss errors thereby mitigating the deterioration of the original quantum information. However, it is conceivable that further errors are introduced due to imperfect gates or to additional environmental effects, such as those occurring when applying logical operations to the GKP logical qubit. One possible way to simulate these errors and test the efficiency of the QEC protocol is to inject displacement errors into the oscillator state, as in \cite{Sivak2023}. In this case, the original logical state (we have chosen $|+ Z_L \rangle$ as a prototypical example) is displaced by a displacement operator $\hat{D}(\alpha)$ as described in Sec.~\ref{logical_ops}, resulting in the perturbed state $\hat{D}(\alpha) | + Z_L \rangle = | + \tilde{Z}_L \rangle$. Due to the structure of the square GKP code, a displacement larger than $\alpha = \sqrt{\pi /2} \approx 0.625$ corresponds to a deterministic logical error. This causes the initial logical state $| + Z_L \rangle$ to flip to $| - Z_L \rangle$, and the former logical information can no longer be recovered. \\
\indent Therefore, by applying the quantum error correction cycles to the shifted state, in the presence of additional photon losses and transmon decoherence, it is possible to track the combined ability of the protocol to recover the original logical information and prevent further degradation. In our numerical simulations we considered the application of real displacements with amplitudes ranging from $\alpha = 0.1$ to $\alpha = 0.6$ (which is the highest value that allows the recovery of the initial logical information) on the initial unperturbed state $| + Z_L \rangle$. The results shown in Fig.~3(c) of the main text illustrate the difference of the logical information of the GKP code when we deploy our non-Markovian protocol and the standard one over the time of applying many correction cycles, namely $\langle \hat{Z} \rangle (\text{NMF}) - \langle \hat{Z} \rangle (\text{std})$. The positive numbers of this set demonstrate the superior capability of our non-Markovian feedback strategy to correct logical errors and recover original quantum information in comparison to the standard strategy. This challenge extends beyond the simple quantum memory task, as additional errors to those inherent in the system can also be corrected.

\subsection{Biased-noise gates QEC}
\begin{figure}[h!]
\centering
\includegraphics[width=12cm]{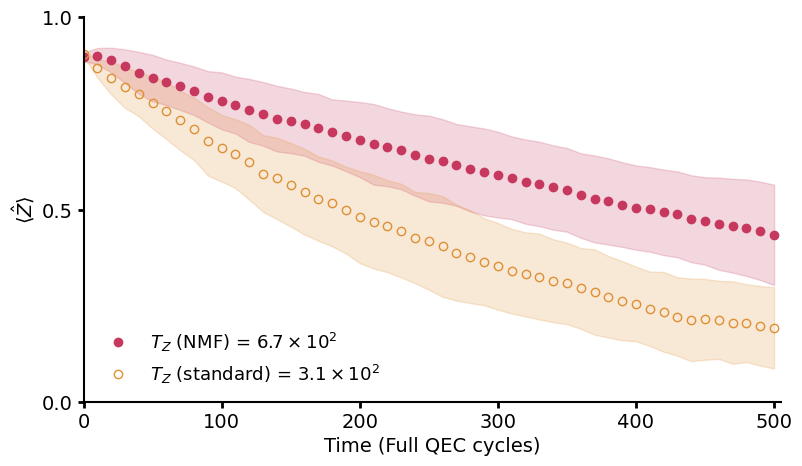}
\caption{QEC with biased gates. Comparison of the QEC performance between the standard (empty dots, orange) and the NMF (full dots, red) protocol using gates with biased errors as in Table \ref{table_err}. The shaded area represents the standard deviation over all the sampled trajectories.} \label{Z_bias}
\end{figure}

We show here that the optimized memory-feedback QEC strategy for the sBs protocol used so far modeling perfect gates, can be generalized also to the case of gates with biased errors constant in time.
In particular, we choose the random biased errors for the gates' parameters as shown in Table \ref{table_err}.\\
\begin{table}[h!]
\begin{tabular}{l*{3}{ccc}}
\hline
Gate & Parameter  & Biased error \\
\hline
$\hat{R}_q$ & $\phi_1$ & + 0.05 \\
$\hat{R}_q$ & $\phi_2$ & - 0.03 \\
$\hat{R}_q$ & $\phi_3$ & - 0.06 \\
$\hat{R}_q$ & $\phi_4$ & + 0.04 \\
$\hat{R}_q$ & $\theta_1$ & - 0.03 \\
$\hat{R}_q$ & $\theta_2$ & + 0.05 \\
$\hat{R}_q$ & $\theta_3$ & + 0.06 \\
$\hat{R}_q$ & $\theta_4$ & +0.04 \\
$EC\hat{D}_{qc}$ & $\beta_1$ & + 0.06 - 0.04 i \\
$EC\hat{D}_{qc}$ & $\beta_2$ & + 0.04 - 0.02 i \\
$EC\hat{D}_{qc}$ & $\beta_3$ & + 0.04 - 0.05 i \\ \hline
\end{tabular}
\caption{Biased errors on gates' parameters. List of the biased errors for each parameter of the gates used in the circuit in Fig.~\ref{scheme_QEC} for the test of biased noise gates. }  \label{table_err}
\end{table}
Comparing the Pauli expectation value $\langle \hat{Z} \rangle$ as a function of time for the optimized and the standard sBs protocol, given the initial state $| + Z_L \rangle$ and applying periodically the QEC cycles, we notice that the lifetime decreases with respect to the case of perfect gates, as expected (Fig.~\ref{Z_bias}). However, the optimized strategy has a lifetime $T_Z (\text{NMF})/ \tau_{cycle} = 3.3 \times 10^2$, while the standard one is $T_Z (\text{standard})/ \tau_{cycle} = 1.6 \times 10^2$, i.e. a factor 2 smaller, similarly to the case of perfect gates.

\subsection{Effect of cavity dephasing} \label{Dephasing}
As mentioned in the supplementary section \ref{noise}, pure white-noise dephasing is typically negligible on the photon number scale of the GKP state. However, in addition to the cavity white-noise dephasing, it is known that additional pure dephasing can also arise from off-resonant coupling of the cavity to the ancillary transmon used to control the cavity \cite{Puri2021, Sivak2023, goldblatt2024}, caused by fluctuations in the transmon population due to its finite temperature (here called \textit{shot-noise dephasing}, consistent with the cited references). The corresponding additional pure dephasing time can reach values around $T_{\phi,c}^{\text{shot}}$ = 5 ms for a free evolution of a superposition of Fock states $| 0 \rangle$ and $| 1 \rangle$ in the cavity coupled to the fluctuating qubit \cite{Sivak2023}. Since the dephasing rate scales with the number of photons, this value alone would imply a significant degradation of the coherence time when considering the typical photon numbers of GKP states (photon numbers of the order of 5). However, the actually observed lifetimes in the experiment \cite{CampagneIbarcq2020, Sivak2023}, even without our improved protocol, are much better than what is suggested by this number. The reason for this is that the transmon is reset to the ground state every cycle and that the relevant gates are implemented with built-in decoupling sequences. This mitigates the dephasing arising at this level.\\
\indent As described in section \ref{sBsprotocol} of this Supplemental Material, we simulate the evolution of the studied system as formed by instantaneous perfect unitary gates followed by an idling time (with a duration equal to the gates’ pulses) during which we let the noise act freely, including the cavity decay and the transmon decay and dephasing. This approach allows to capture at least in a time-averaged way the physics of the noisy evolution of our system, resulting in imperfect gate realizations. 
\begin{figure}[h!]
\centering
\includegraphics[width=10cm]{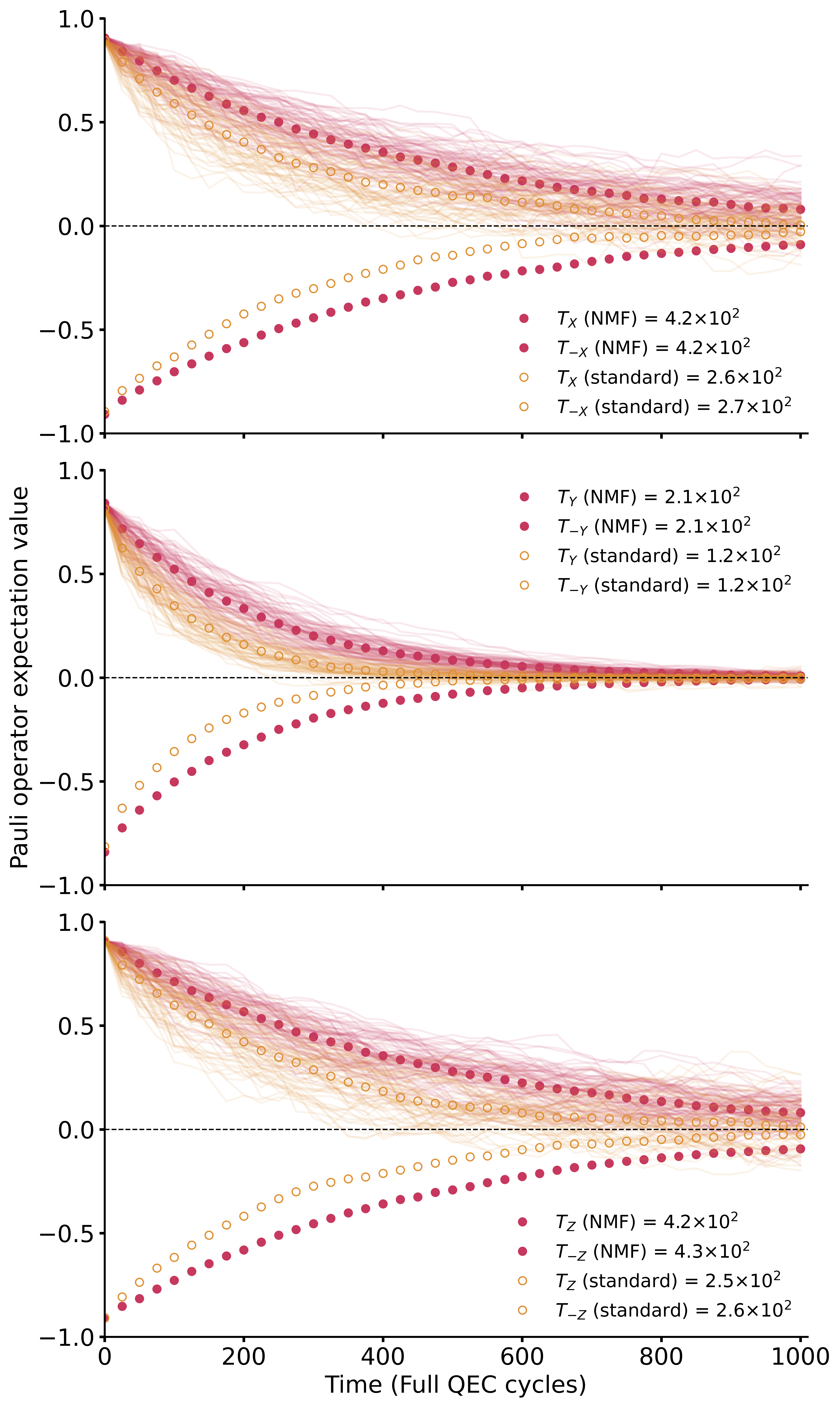}
\caption{Time evolution of logical states including cavity dephasing. Comparison of the logical state $\lvert \pm X_L\rangle$ (top), $\lvert \pm Y_L\rangle$ (center) and $\lvert \pm Z_L\rangle$ (bottom) time evolution (dots: average over multiple trajectories, continuous faded lines: sampled trajectories averaged over one batch of trajectories) for the standard sBs protocol (yellow) and the NMF approach (red) advocated in this work. The calculations have been performing adding a cavity dephasing time $T_{\phi,c}^{\text{lump}} = 24$ ms, as explained in the text, without retraining of the neural network.} \label{Pauli_dephasing}
\end{figure}
To extend our initial simulations and consider the impact of dephasing, we propose a methodology wherein all residual dephasing is lumped into a standard white-noise Lindblad term, which can be readily integrated into our model. As a ballpark figure for the additional dephasing time, we adopted a value of $T_{\phi,c}^{\text{lump}} = 24$ ms, which is the scale of the unexplained residual additional dephasing extracted in the analysis of experimental data (see section 2E in the Supplementary Material of Ref. \cite{Sivak2023}). 
Therefore, supplementary simulations were performed adopting this picture, comparing the standard and our non-Markovian strategy, without retraining of the neural network. The results of the time evolution of the different logical states ($| \pm X_L \rangle$,$| \pm Y_L \rangle$ and $| \pm Z_L \rangle$) are shown in Fig.~\ref{Pauli_dephasing}. The lifetime of the GKP code when adopting the standard quantum error correction strategy is $T_{Z_L}(\text{std}) = 2.1$ ms, while with our non-Markovian feedback approach we get $T_{Z_L} (\text{NMF}) = 3.5$ ms. It is noteworthy that even in the presence of additional noise and with no further optimization of the gates’ parameters, the non-Markovian strategy continues to outperform the standard one by a factor of approximately $\approx 1.7$, thereby confirming the robustness of our results.